\documentclass[%
 reprint,
 superscriptaddress,
 showkeys,
nofootinbib,
 amsmath,amssymb,
 aps,
prc,
]{revtex4-2}
\usepackage{longtable}
\usepackage{pgfplotstable}
\usepackage{graphicx}
\usepackage{dcolumn}
\usepackage{bm}
\usepackage{float}
\usepackage{xcolor}
\usepackage{hyperref}
\usepackage{tabularx}
\usepackage{booktabs}
\usepackage{bigdelim}
\usepackage{afterpage}

\DeclareUnicodeCharacter{2212}{-}

\definecolor{lightgray}{rgb}{0.77, 0.76, 0.82}

\newcommand{\asymuncert}[3]{${#1}^{+{#2}}_{-{#3}}$}

\newcolumntype{H}{>{\setbox0=\hbox\bgroup}c<{\egroup}@{}}
\newcolumntype{L}[1]{>{\raggedright\let\newline\\\arraybackslash\hspace{0pt}}m{#1}}
\newcolumntype{C}[1]{>{\centering\let\newline\\\arraybackslash\hspace{0pt}}m{#1}}
\newcolumntype{R}[1]{>{\raggedleft\let\newline\\\arraybackslash\hspace{0pt}}m{#1}}

\begin{document}


\title{Electric and magnetic dipole strength in $^{58}$Ni from forward-angle proton scattering}

\author{I. Brandherm}
\affiliation{Institut {f\"u}r Kernphysik, Technische Universit{\"a}t Darmstadt, D-64289 Darmstadt, Germany}

\author{P.~von Neumann-Cosel}\email{email:vnc@ikp.tu-darmstadt.de}
\affiliation{Institut {f\"u}r Kernphysik, Technische Universit{\"a}t Darmstadt, D-64289 Darmstadt, Germany}

\author{R.~Mancino}
\affiliation{Institut {f\"u}r Kernphysik, Technische Universit{\"a}t Darmstadt, D-64289 Darmstadt, Germany}
\affiliation{GSI Helmholzzentrum f\"ur Schwerionenforschung, Planckstraße 1, 64291 Darmstadt, Germany}
\affiliation{Faculty of Mathematics and Physics, Charles University, Prague, Czech Republic}

\author{G.~Mart\'inez-Pinedo}
\affiliation{Institut {f\"u}r Kernphysik, Technische Universit{\"a}t Darmstadt, D-64289 Darmstadt, Germany}
\affiliation{GSI Helmholzzentrum f\"ur Schwerionenforschung, Planckstraße 1, 64291 Darmstadt, Germany}

\author{H.~Matsubara}
\affiliation{Research Center for Nuclear Physics, Osaka University, Ibaraki, Osaka 567-0047, Japan}
\affiliation{Faculty of Radiological Technology, Fujita Health University, Aichi 470-1192, Japan}

\author{V.~Yu.~Ponomarev}
\affiliation{Institut {f\"u}r Kernphysik, Technische Universit{\"a}t Darmstadt, D-64289 Darmstadt, Germany}

\author{A.~Richter}
\affiliation{Institut {f\"u}r Kernphysik, Technische Universit{\"a}t Darmstadt, D-64289 Darmstadt, Germany}

\author{M.~Scheck}
\affiliation{School of Computing, Engineering, and Physical Sciences, University of the West of Scotland, Paisley, PA1 2BE, United Kingdom}

\affiliation{SUPA, Scottish Universities Physics Alliance, United Kingdom} 


\author{A.~Tamii}
\affiliation{Research Center for Nuclear Physics, Osaka University, Ibaraki, Osaka 567-0047, Japan}

\date{\today}

\begin{abstract}
\begin{description}
\item[Background]
Electric and magnetic dipole strengths in nuclei at excitation energies well below the giant resonance region are of interest for a variety of nuclear structure problems including a possible electric dipole toroidal mode or the quenching of spin-isospinflip modes.

\item[Purpose]
The aim of the present work is a state-by-state analysis of possible $E1$ and $M1$ transitions in $^{58}$Ni with a high-resolution $(p,p^\prime)$ experiment at 295 MeV and very forward angles including $0^\circ$ and a comparison to results from studies of the dipole strength with the $(\gamma,\gamma^\prime)$ and $(e,e^\prime)$ reactions.

\item[Methods]
The $E1$ and $M1$ cross sections of individual peaks in the spectra are deduced with a multipole decomposition analysis (MDA). 
They are converted to reduced $E1$ and spin-$M1$ transition strengths using the virtual photon method of relativistic Coulomb excitation and the unit cross-section method, respectively. 
The experimental $M1$ strength distribution is compared to large-scale shell-model calculations with the effective GXPF1A and KB3G interactions.

\item[Results]
In total, 11 $E1$ and 26 $M1$ transitions could be uniquely identified in the excitation energy region $6 - 13$~MeV.
In addition, 22 dipole transitions with preference for either $E1$ or $M1$ multipolarity and 57 transitions with uncertain multipolarity were found.
Despite the high level density good agreement is obtained for the deduced excitation energies of $J = 1$ states in the three types of experiments indicating that the same states are excited. 
The $B(E1)$ and $B(M1)$ strengths deduced in the $(\gamma,\gamma^\prime)$ experiments are systematically smaller than in the present work because of the lack of information on branching ratios to lower-lying excited states and the competition of particle emission.
Fair agreement with the $B(M1)$ strengths extracted from the $(e,e^\prime)$ data is obtained after removal of $E1$ transitions uniquely assigned in the present work belonging to a low-energy toroidal mode with unusual properties mimicking $M1$ excitations in electron scattering.
The shell-model calculations provide a good description of the isospin splitting and the running sum of the $M1$ strength.
A quenching factor 0.74 for the spin-isospin part of the $M1$ operator is needed to attain quantitative agreement with the data.

\item[Conclusions]
High-resolution forward-angle inelastic proton scattering experiments at beam energies of about 300 MeV are a highly selective tool for an extraction of resolved $E1$ and $M1$ strength distributions in medium-mass nuclei.
Fair agreement with results from electron scattering experiments is obtained indicating a dominance of spin contributions to the $M1$ strength. 
Shell-model calculations are in good agreement with gross properties of the $M1$ strength distribution when a quenching factor for the spin-isospin part comparable to the one needed for a description of Gamow-Teller (GT) strength is included. 

\end{description}
\end{abstract}


\maketitle


\section{\label{sec:Intro}Introduction}

Magnetic dipole excitations represent an elementary mode of low-energy nuclear structure.
The transition strengths contain coherent contributions from spin and orbital currents.
In heavy nuclei transitions with dominant spin or orbital contributions are energetically well separated and form the orbital scissors mode and a spinflip resonance with energy centroids of about $15 \times A^{-1/3}$ and $40 \times A^{-1/3}$, respectively \cite{hey10}.
In lighter nuclei the two contributions are mixed with sizable interference effects in individual transitions (see, e.g., Refs.~\cite{fuj97,hof02}).

Because of the anomalous proton and neutron magnetic moments, isovector (IV) strength strongly dominates over isoscalar (IS) strength and spinflip strengths are enhanced with respect to orbital strengths \cite{lan04}.
Since the structure of the $M1$ operator does not allow for changes of the radial quantum number, in a shell-model picture $M1$ transitions are restricted to spin-orbit partners.
Thus, systematic investigations of the spinflip $M1$ strength can provide insight into shell evolution driven by the tensor force \cite{ots20}.
Knowledge of the spin $M1$ strength distributions is also important for the description of neutral current neutrino reactions in astrophysical environments \cite{lan04,lan10} and possible signals in future underground neutrino detectors \cite{gay19,tor22}.

Assuming isospin symmetry, IV spinflip $M1$ transitions are analogs of Gamow-Teller (GT) transitions with isospin $T = T_0$, where $T_0$ denotes the isospin of the target gound state (g.s.).
For $T_0 \neq 0$, selection rules also allow the excitation of states with $T_0 + 1$, i.e., the analog of $\mathrm{GT}_+$ transitions \cite{fuj11}.
Thus, spinflip $M1$ strength distributions are an independent way to investigate the long-standing problem of quenching of the axial strength in nuclei. 
In $fp$-shell nuclei a comparable reduction factor for GT $\beta$ decay \cite{mar96} and total $B(M1)$ \cite{vnc98} strengths has been established in comparison with shell-model calculations providing insight into the dominant quenching mechanisms.

It has been shown that a significant contribution to GT quenching arises from two-body currents and can be quantitatively described by coupled-cluster \cite{gys19} and shell-model \cite{cor24} calculations based on chiral effective field theory ($\chi$EFT) interactions.
The recent development of a corresponding $\chi$EFT framework \cite{seu23} for shell-model calculations of the magnetic dipole strength makes new experimental investigations particularly interesting.

The present work reports on spinflip $M1$ strength in $^{58}$Ni extracted from high-resolution proton inelastic scattering at extreme forward angles including $0^\circ$.
The nucleus $^{58}$Ni has been subject of extensive studies of the $M1$ strength distribution with photon \cite{bau00,sch13,shi24} and electron \cite{met87} scattering as well as GT$_\pm$ strength with high-resolution charge-exchange reactions \cite{fuj07,hag05}.
The new data shed light on discrepancies in the parity of assignments for some transitions between the $(e,e^\prime)$ and $(\gamma,\gamma^\prime)$ experiments, where $M1$ character was assigned in the former and $E1$ in the latter.
As discussed elsewhere \cite{vnc23}, these are toroidal $E1$ excitations \cite{rep19} with unusual properties mimicking the signatures of $M1$ transitions in low-momentum transfer electron scattering. 

A pioneering $0^\circ$ inelastic proton scattering experiment on $^{58}$Ni was performed at the Indiana University Cyclotron Facility (IUCF) and used in Ref.~\cite{fuj07} to determine final-state isospins by comparison of relative cross section to charge-exchange reactions \cite{hag05,fuj02}. 
However, the experiment was limited to $0^\circ$ measurements only and no independent determination of the multipolarity based on the angular distributions was possible.
Furthermore, the data contained a large instrumental background limiting the sensitivity.
The results reported here obtained at the $0^\circ$ facility at the Research Center for Nuclear Physics (RCNP), Osaka, Japan \cite{vnc19}, are based on background-subtracted spectra (obtained with the methods described in Ref.~\cite{tam09}) and includes measurements at finite angles which permit a MDA.
The excellent energy resolution of the newly reported data enables a state-by-state analysis analog to the study of spin-$M1$ strength in $^{48}$Ca \cite{mat17} up to excitation energies of about 13.5 MeV.

The paper is organized as follows.
Section \ref{sec:experiment} describes the experiment and the resulting spectra.
Section \ref{sec:analysis} provides information on the data analysis, details of the MDA, and the methods of conversion from cross sections to transition strengths.
Results and a detailed comparison with previous work are presented in Sec.~\ref{sec:results}.
Shell-model calculations of the $B(M1)$ and spin-$M1$ strength distributions are discussed in Sec.~\ref{sec:shell model}.
Finally, Sec.~\ref{sec:conclusion} provides a summary. 

\section{\label{sec:experiment}Experiment}

\subsection{Details of the experiment at RCNP}

The experiment was performed at the RCNP cyclotron facility. 
An unpolarized proton beam with an energy of 295 MeV impinged on a highly enriched ($>$98\% $^{58}$Ni) nickel target with an areal density of 4 mg/cm$^2$. 
Typical proton beam currents during the experiment were $3 - 7$ nA. 
Scattered protons were detected with the Grand Raiden (GR) magnetic spectrometer \cite{fuj99}. 
The GR provides a special setup for $0^\circ$ scattering experiments \cite{tam09}. 
Measurements were performed with the GR placed at angles of $0^\circ$, $2.5^\circ$, and $4.5^\circ$.
A detailed description of the raw data analysis can be found in Ref.~\cite{tam09}.  

The large angular acceptance of the GR spectrometer enables the extraction of several spectra with different scattering angles from one spectrometer setting. 
In doing so spectra for laboratory scattering angles $0.40^\circ$, $1.00^\circ$, $1.74^\circ$ (spectrometer angle $0^\circ$), $2.38^\circ$, $3.18^\circ$ (spectrometer angle $2.5^\circ$), and $4.39^\circ$, $5.15^\circ$ (spectrometer angle $4.5^\circ$) were obtained. 
An excellent energy resolution of 22 keV full width at half maximum (FWHM) was achieved by applying dispersion matching between the incoming proton beam and the spectrometer. 
Additionally, energy calibration measurements with a $^{12}$C target were performed. 
A detailed description of the raw data analysis can be found in \cite{tam09}. 

The experimental uncertainties of the double differential cross sections are summarized in Table~\ref{tab:error_contributions} and include solid angle determination, charge collection, uncertainty in target thickness and enrichment, drift chamber efficiency, and data acquisition life time ratio. 
Assuming independence of the individual contributions, the total systematic uncertainty amounts to 6.5\%. 
Statistical errors are generally much smaller due to the high statistics. 

\begin{table}[h]
\caption{\label{tab:error_contributions} Error contributions to the differential cross sections.}
\begin{tabular}{lc}
\hline
\hline
Solid angle & 5\% \\
Charge collection  & 3\% \\
Target thickness & 1.5\% \\
Live time ratio & 1.5\% \\
Detector efﬁciency  & 1.7\% \\
\hline
Total & 6.5\% \\
\hline
\hline
\end{tabular}
\end{table}

\subsection{Spectra}

\begin{figure}
    \centering
    \includegraphics[width = 0.9\columnwidth]{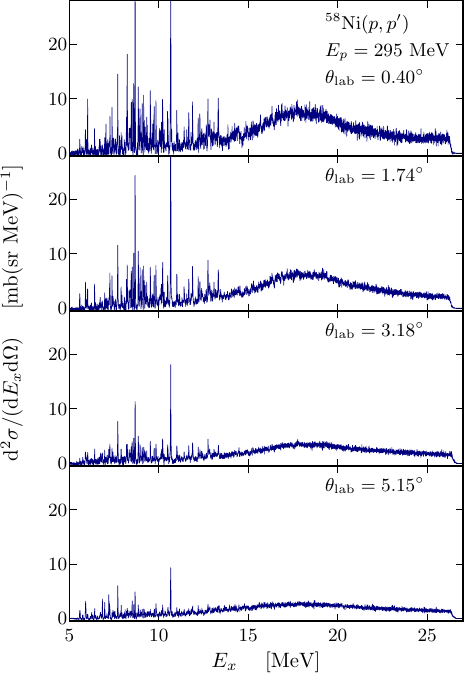}
    \caption{Excitation energy spectra of the $^{58}$Ni$(p,p^\prime)$ reaction measured at the indicated scattering angles. 
    }
    \label{fig:ni_spectra}
\end{figure}

\begin{figure*}
\centering
{
\includegraphics[scale=1]{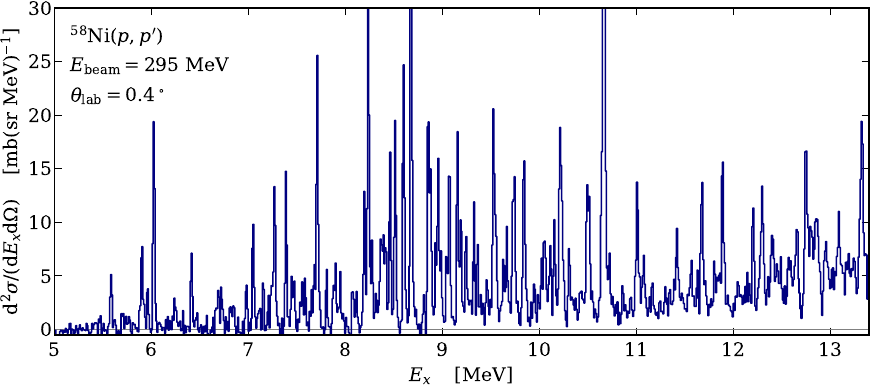}}
\caption{Excitation energy spectrum of the $^{58}$Ni$(p,p^\prime)$ reaction between 5 and 13 MeV at a scattering angle $\theta_{\rm lab} = 0.4^\circ$.} 
\label{fig:ni58_04deg}
\end{figure*}

Figure \ref{fig:ni_spectra} shows examples of the $^{58}$Ni$(p,p^\prime)$ spectra for scattering angles $0.40^\circ$, $1.74^\circ$, $3.18^\circ$, and $5.15^\circ$. 
They cover an excitation energy range from 5 MeV to 26 MeV. 
A large number of well resolved lines can be seen up to an excitation energy of approximately 13 MeV, each corresponding to an excited state. 
For the vast majority of states, the differential cross section drops with increasing scattering angle indicating a dipole character. 
The broad resonance-shaped structure centered at about 18.5 MeV is the isovector electric giant dipole resonance (IVGDR) most pronounced at small scattering angles. 

A state-by-state analysis of the peaks in the spectra is presented in the following. 
A detailed view of the region of interest for the spectrum taken at a scattering angle $0.4^\circ$ is presented in Fig. \ref{fig:ni58_04deg}.
The scattering around the zero line, which is somewhat vertically shifted, after background subtraction demonstrates that the spectrum is essentially background free below $\approx 9$\,MeV. 
A slowly increasing background appears towards higher excitation energies. 
This is mainly caused by unresolved strength due to the level density increasing with excitation energy and the emerging tail of the IVGDR. 
Since the proton separation threshold lies at $S_p = 8.172$ MeV, one cannot exclude small contributions from quasi-free scattering although they should be suppressed by the Coulomb barrier up to energies of about 2 MeV above $S_p$ . 
Above the neutron separation threshold $S_n = 12.216$ MeV a background contribution caused by quasi-free scattering is expected.
However, earlier $(p,p^\prime)$ experiments with the same technique show a slow rise with excitation energy \cite{bir17,bas20,fea23} indicating small contributions in the region considered here. 

The excitation energies of the spectra were locally recalibrated using a quadratic function. 
For this purpose the eight most pronounced lines in the $^{58}$Ni$(p,p')$ spectra were used, for which the excitation energies are precisely determined in $(\gamma,\gamma^\prime)$ experiments \cite{bau00,sch13,shi24}. 
With this approach an uncertainty of $\pm10$~keV is achieved.

\section{\label{sec:analysis}Data Analysis}

\subsection{\label{sec:state-by-state}State-by-state analysis}

\begin{figure}[b]
    \centering
    \includegraphics[width = 0.9\columnwidth]{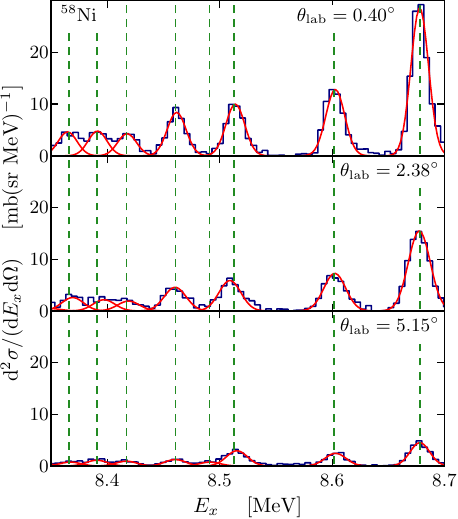}
    \caption{Spectra of the $^{58}$Ni$(p,p')$ reaction in the excitation energy range of $E_x = 8.35 - 8.7$ MeV at scattering angles $\theta_\mathrm{lab} = 0.4^\circ$, $2.68^\circ$, and $5.15^\circ$. 
    Red curves are Gaussians, fitted to the peaks. 
    Dashed green lines mark the average centroid energies.}
    \label{fig:ni58_peaks}
\end{figure}

In the following, we present a state-by-state analysis.
Excitation energy regions of a several hundred keV typically containing 10 to 20 peaks were fitted simultaneously with Gaussians of equal width allowing for a piecewise linear background.
The software HDTV was used \cite{HDTV}. 
With the exception of a few states at the highest excitation energies, no variations in line widths or shapes were observed. 
The high level density complicates an unambiguous determination of the height and shape of the background, since background-only regions in the spectra between the peaks are scarce. 

Next, the consistent appearance of peaks in spectra at different scattering angles was investigated as illustrated in Fig.~\ref{fig:ni58_peaks}.
In total 147 peaks were found with the constraint that they appear in at least 5 of the 7 spectra available. 


%

\subsection{\label{subsec:mda}Multipole decomposition analysis}

\begin{figure}[b]
    \centering
    \includegraphics[width = 0.9\columnwidth]{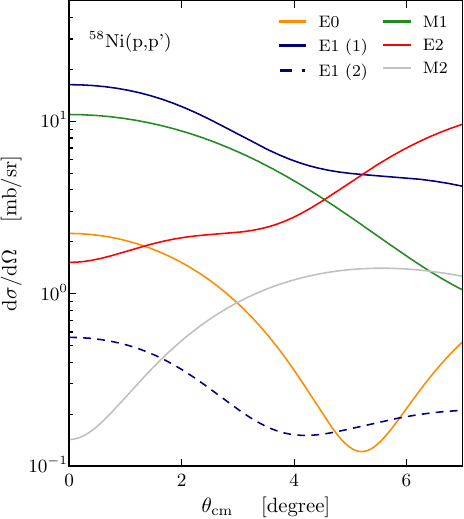}
    \caption{Theoretical angular distributions for transitions of different multipolarities in $^{58}$Ni$(p,p^\prime)$ reaction, calculated with the computer code DWBA07 \cite{ray07} for an incident proton beam energy of 295 MeV. 
    Two different curves for $E1$ transitions are used in the analysis labeled (1) and (2).}
    \label{fig:DWBA}   
\end{figure}

\begin{figure}
    \centering
    \includegraphics[width =\columnwidth]{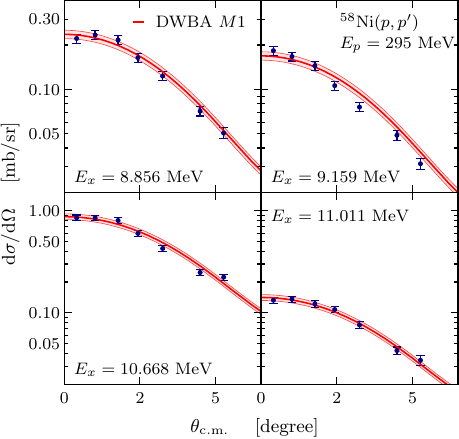}
    \caption{Experimental angular distributions of the $^{58}$Ni$(p,p^\prime)$ reaction for transitions assigned multipolarity $M1$. 
    The red solid lines are the theoretical predictions with uncertainty bands from a Monte Carlo variation of experimental uncertainties.}
    \label{fig:m1_angdist}
\end{figure}

\begin{figure}[b]
    \centering
    \includegraphics[width =\columnwidth]{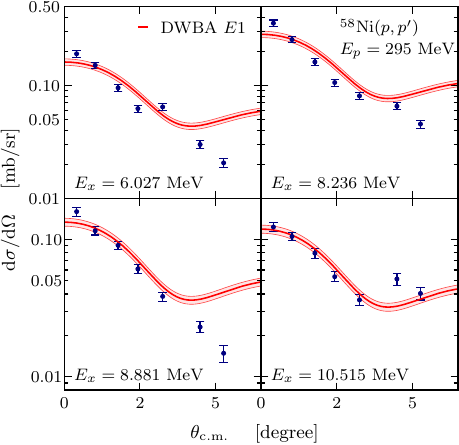}
    \caption{Same as Fig.~\ref{fig:m1_angdist}, but for $E1$ transitions.}
    \label{fig:e1_angdist}
\end{figure}

Information about spin and parity of an excited state is obtained from the angular distribution of the scattered protons. 
Theoretical angular distributions for transitions with different multipolarities were calculated for the $^{58}$Ni$(p,p^\prime)$ reaction in distorted wave Born approximation (DWBA) with the computer code DWBA07 \cite{ray07} using the Love and Franey effective proton-nucleus interaction \cite{lov81}. 
Transition amplitudes and single-particle wave functions from quasi-particle phonon model (QPM) calculations similar to Refs.~\cite{bir17,fea23} served as input. 

The angular distributions used in this work are shown in Fig.~\ref{fig:DWBA}.
They represent the most collective excitation for each multipolarity.
Several angular distributions were considered for $E1$ transitions to take into account effects of Coulomb-nuclear interference. 
They showed a different behaviour at low and IVGDR energies, but the resulting angular distributions were almost identical for transitions in either energy region.
Thus, only one curve representative for each energy region was used in the following, cf.~Fig.~\ref{fig:DWBA}.

The $E0$ response in $^{58}$Ni was studied with the $(\alpha, \alpha')$ reaction by Lui {\it et al.} \cite{lui06}, where it was shown that 
the largest part of the $E0$ strength resides at excitation energies well above the region investigated here.
Furthermore, monopole transitions are only weakly excited in the $(p,p^\prime)$ reaction.
The $E0$ transition plotted in Fig.~\ref{fig:DWBA} contains a significant fraction of the isoscalar giant monopole resonance strength but is still about an order of magnitude smaller at $0^\circ$ than strong $M1$ and $E1$ excitations. 
Thus, monopole excitations were neglected in the MDA. 
Then, experimental angular distributions with a clear maximum at $0^\circ$ are a signature for dipole transitions.

With the exception of $E2$, transitions with multipolarities $\lambda > 1$ show a strong decrease of the angular distributions towards $\theta_{\rm cm} = 0^\circ$ with a very similar slope in the forward-angle range covered by the experiment (cf.~Ref.~\cite{vnc19}).
Thus, they are represented by a single curve ($M2$) in the MDA.
The superposition of at most two unresolved levels with different $J^\pi$ was considered.
A further discussion of the assumptions and approximations  underlying the present MDA approach can be found in Ref.~\cite{vnc19} and the corresponding formulas e.g.\ in Ref.~\cite{mat17}.

Examples of angular distributions for known $M1$ and $E1$ transitions in $^{58}$Ni are shown in Figs.~\ref{fig:m1_angdist} and \ref{fig:e1_angdist}, respectively. 
The uncertainty bands are determined from a Monte Carlo variation of the experimental error contributions summarized in Table~\ref{tab:error_contributions}.
For the $M1$ cases, the theoretical angular distribution shows excellent agreement with the data. 
For the small momentum transfers covered by the experiment, IV spin-$M1$ excitations exhibit an almost universal behavior independent of their particle-hole structure. 

The situation is somewhat more complex for $E1$ transitions.
Generally, they show a steeper decrease than $M1$'s at very small  angles $ \theta_{\rm cm} < 2^\circ$, which serves to distinguish between the two multipolarities.
At angles $2 < \theta_{\rm cm} < 5 ^\circ$, where effects of Coulomb-nuclear interference are significant, larger deviations between experimental and theoretical angular distributions are observed in many cases as illustrated in Fig.~\ref{fig:e1_angdist}.
In particular, these transitions do not show the theoretically predicted minimum around $4^\circ$ but rather a monotonous decrease with scattering angle.  
A possible explanation of these discrepancies could be a different sign of the Coulomb-nuclear interference term (generally predicted to be constructive).
In any case, this leads to some ambiguity in the normalization of the theoretical curves to the data and thus the extracted $B(E1)$ strength.

\begin{figure}
    \centering
    \includegraphics[width =\columnwidth]{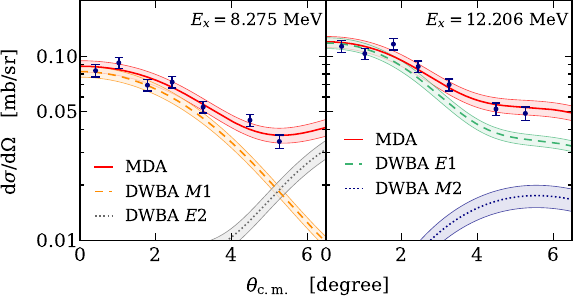}
    \caption{Same as Fig.~\ref{fig:m1_angdist}, but assuming two unresolved transitions with different multipolarity.}
    \label{fig:mixed_angdist}
\end{figure}

Figure \ref{fig:mixed_angdist} presents examples of the MDA where assuming an unresolved doublet of states with different $J^\pi$ provides the best description of the data. 

\subsection{Conversion to transition strengths}
\label{subsec:conversion}

The $M1$ and $E1$ cross sections obtained from the MDA can be converted to transition strengths with the unit cross section method for the former and assuming relativistic Coulomb excitation for the latter.

The unit cross section method was developed to extract the reduced Gamow-Teller transition strength $B(\mathrm{GT})$ from the cross section of a charge-exchange (CE) reaction at $0^\circ$ \cite{tad87}. 
At beam energies $> 100$ MeV/nucleon one-step processes dominate CE cross sections.
Then, measured cross section and the reduced transition matrix element can be related by a constant factor called unit-cross section $\hat{\sigma}$. 
At $0^\circ$, cross sections and transition strengths are related by
\begin{equation}
\label{eq:unit_crs}
\frac{\mathrm{d}\sigma}{\mathrm{d}\Omega}\left(0^\circ\right)=\hat{\sigma}_\mathrm{GT}\cdot F(q,E_{\rm x})\cdot B(\mathrm{GT}),
\end{equation}
where $\hat{\sigma}_\mathrm{GT}$ depends on the nuclear mass and the projectile energy of the reaction. 
Kinematical corrections for non-zero momentum transfer are contained in a correction factor $F(q, E_{\rm x})$, which can be obtained by DWBA calculations. 
In Ref.~\cite{sas09} $(p,n)$ charge-exchange reactions at the same incident energy as in the present $(p,p^\prime)$  experiment have been evaluated, and a mass-dependent expression for $\hat{\sigma}_\mathrm{GT}$ was derived
\begin{equation}
	\hat{\sigma}_\mathrm{GT} = 3.4(3) \cdot \exp \left[-0.40(5)\left(A^{1/3}-90^{1/3}\right)\right]
\end{equation}
resulting in $\hat\sigma_\mathrm{GT} = 4.34$ mb/sr for $^{58}$Ni at an incident proton energy of 295 MeV. 

It has been shown, that this method can be extended to the extraction of the isovector spin-flip $M1$ transition strength \cite{bir16} utilizing isospin symmetry \cite{fuj11} and neglecting small isoscalar contributions to the $(p,p^\prime)$ cross sections.
For small momentum transfers $\hat{\sigma}_\mathrm{GT} \simeq \hat{\sigma}_\mathrm{M1}$. 
The resulting IV spin-flip $M1$ strength is denoted $B({\rm GT}_0)$ in the following because the excited states are isospin analogs of the $B({\rm GT}_\pm)$ transitions with $T_f = T_0$, where $T_0$ denotes the ground-state isospin.
However, one should note that the selection rules also permit the excitation of states with $T_f = T_0 + 1$. 

For a direct comparison to experiments using electromagnetic probes, like $(\gamma,\gamma')$ and $(e,e')$, it is convenient to convert the $B({\rm GT}_0)$ values to the equivalent electromagnetic reduced transition strength $B(M1)$. 
The latter is defined for a transition from initial state $i$ to final state $f$ as
\begin{equation}
    B(M1) = \frac{3}{4\pi} \frac{1}{2J_i+1}\bigl|\langle f||\vec{\mu} ||i \rangle\bigr|\mu_N^2,
\end{equation}
where $\vec{\mu}$ denotes the magnetic dipole operator. 
The magnetic dipole operator can be decomposed into isoscalar (IS) and isovector (IV) terms
\begin{align}
    B(M1) = &\frac{3}{4\pi} \frac{1}{2J_i +1} \bigl|\langle f||g_l^\mathrm{IS}\vec{l} + 
    \frac{g_s^\mathrm{IS}}{2}\vec{\sigma} + \\ \nonumber
    &\left( g_l^\mathrm{IV}\vec{l} + \frac{g_s^\mathrm{IV}}{2}\vec{\sigma}\right)\tau_0 || i \rangle \bigr|\mu_N^2
\end{align}
with $g$-factors $g_s^\mathrm{IS}=g_s^\mathrm{IV} = 0.5$, $g_l^\mathrm{IS} = 0.88$ and $g_l^\mathrm{IV}=4.706$. 
Spin-Pauli matrices and orbital angular momentum are denoted as $\vec{\sigma}$ and $\vec{l}$ respectively, and the isospin operator as $\tau_0$. 
In the extraction of $M1$ strengths from the $(p,p^\prime)$ reaction terms containing the angular momentum operator as well as isoscalar contributions are neglected leading to 
\begin{equation}
    \label{eq:em_BM1}
	B(M1) = \frac{3}{2\pi}\left(\frac{g_s^{\mathrm{IV}}}{2}\right)^2 B(\mathrm{GT}_0) \,\, \mu_N^2. 
\end{equation}
A detailed discussion of the underlying assumptions and related systematic uncertainties can be found in Refs.~\cite{bir16,fuj11}.

Conversion of $E1$ cross sections to reduced $B(E1)$ transition strengths is achieved with the aid of the virtual photon method relating relativistic Coulomb excitation and equivalent photoabsorption cross sections $\sigma_\gamma$ \cite{ber88}
\begin{equation}
\label{eq:virtual photon method}
\frac{{\rm d}^2 \sigma}{{\rm d}\Omega{\rm d}E_\gamma} = \frac{1}{E_\gamma} \frac{{\rm d} N_{E1}}{{\rm d}\Omega} \sigma_\gamma(E_\gamma).
\end{equation}
Here, $N_{E1}$ is the number of virtual photons with multipolarity $E1$ calculated in the present work in an eikonal approach \cite{ber93}.
For examples of the energy and angular dependence of $N_{E1}$ in the present kinematics, see Ref.~\cite{vnc19}. 
The method has been verified for discrete transitions in a study of $^{208}$Pb \cite{pol12}.

\section{\label{sec:results} E1 and M1 strength distributions}

\subsection{Results and comparison with other experiments}
\label{subsec:comparison_general}

\afterpage{
\scriptsize
\def\arraystretch{1.25}
	\begin{longtable*}{ccccccccccccc}
     \caption{\label{tab:all} 
     Dipole transitions observed in the present experiment and comparison with results from 
     $(e,e^\prime)$ \cite{met87} and $(\gamma,\gamma^\prime)$ \cite{bau00,sch13,shi24} experiments.} \\
		\toprule
		\toprule
		\multicolumn{4}{c}{$(p,p')$} & \multicolumn{3}{c}{$(e,e')$ } & \multicolumn{6}{c}{$(\gamma,\gamma')$ }\\
		\cmidrule(lr{0.75em}){1-4}\cmidrule(lr{0.75em}){5-7}\cmidrule(lr{0.75em}){8-13}
		$E_{\rm x}$ & $J^\pi$&$B(M1)\!\uparrow$ & $B(E1)\!\uparrow$ &$E_{\rm x}$ &  $J^\pi$& $B(M1)\!\uparrow$ & $E_{\rm x}$\footnotemark[2]  &$J^\pi$\footnotemark[2]&$J^\pi$\footnotemark[3]&$J^\pi$\footnotemark[4]& $B(M1)\!\uparrow$\footnotemark[4] & $B(E1)\!\uparrow$\footnotemark[4] \\
		$[$Me$V$] &  &  $[\mu_N^2]$& $[10^{-3}\mathrm{fm}^2e^2]$ & $[$Me$V$] &  & $[\mu_N^2]$& $[$Me$V$] & & && $[\mu_N^2]$& $[10^{-3}\mathrm{fm}^2e^2]$\\	
		\midrule
		\endhead
		\bottomrule
		\bottomrule

		\endfoot
 %
 6.027 &  $1^{(-)}$   & & 
  &  6.031 & $(1^-)$, $2^+$ &         &  6.027 & $1$ & $1^-$ & $1^-$ &              & 5.69(29) \\
 %
 %
 6.474 & (1)      & \asymuncert{0.028}{0.004}{0.004}
 &  &  6.475 & $1^+$, $(2^-)$ & 0.17(5) & & & & \\
 %
 %
 7.050 &  $1^-$  & & &  7.051 &                &         &  7.048 & $1^-$ &   $1^-$& $1^-$&           & 4.52(14) \\
 %
 %
 7.155 & $1^{(+)}$ & \asymuncert{0.042}{0.006}{0.005} &  &       &                &         &         &       &           &          \\
 %
 %
 7.254 &  $(1)$     & & &  7.255 & $2^+$          &         &  7.250 & $(1)$ &  $1^-$ && &            0.37(13) \\
 7.273 & $1^-$  & \asymuncert{0.171}{0.022}{0.018} & \asymuncert{10.7}{0.8}{0.8}&  7.290 &                &         &  7.272 & $1$ &  $1^-$ & $1^+$ &     0.31(3)       & 3.40(40)   \\
 7.390 &  $1^{(+)}$     & \asymuncert{0.172}{0.023}{0.019}&  &  7.388 & $1^+$           & 0.33(7) &  7.389 & $1^+$ & $1^+$ & $1^+$ &   0.294(15) &          \\  
 %
 %
 7.471 &  $1$     & \asymuncert{0.047}{0.006}{0.006}&  &  7.470 & $1^+$, $(2 ^-)$& 0.25(5) &         &       &           &          \\
 7.554 & $(1)$       & & &  7.560 & $1^+$          & 0.15(4) &         &       &           &          \\
 \multirow{2}{*}{7.590}&\multirow{2}{*}{(1)}&\multirow{2}{*}{} & & & &&  7.585 & & &  $(1)^{-}$ & &    0.58(27) \\
 & & &  & 7.603 & $(1^-)$   & & 7.596 & $(2)$ &           &          \\
 7.650 &  $1$     & \asymuncert{0.040}{0.005}{0.006}& &        &                &         &   &  &  &  \\
 7.711 &   $(1)$    & \asymuncert{0.35}{0.05}{0.04} & &  7.715 & $1^+$          & 0.74(5) &  7.710 & $1^+$ &  $1^+$  & $1^+$  &  0.358(13) &          \\
 %
 %
 %
 8.076 &  $1$     &  &\asymuncert{4.0}{0.3}{0.4} &        &                &         &  8.069 & $(1)^{(-)}$ &   $1^{-}$ & &        & 1.81(24) \\
 %
 %
 8.197 &$1^+$\footnotemark[1]& \asymuncert{0.165}{0.022}{0.017}& &        &                &         &  8.096 & $1$ & & &  0.139(26) & 1.54(29) \\
 \hspace{0.75ex}8.236\footnotemark[5] &  $1^-$ & & \asymuncert{24.4}{1.6}{1.7}&  8.240 & $1^+$          & 1.27(20)&  8.237 & $1^-$ & $1^-$ & $1^-$ &  & 18.41(28)\\
 8.275 &$1^+$\footnotemark[1]& \asymuncert{0.110}{0.014}{0.013}& &  8.276 & $1^+$, $(2^-)$ & 0.26(3) &         &       &           &          \\
 8.323 &  $(1)$  & & &   &   &   &  8.317 & $1$ & $1^-$ & $1^-$ &   & 1.19(18) \\
 8.366 &$1^+$\footnotemark[1]& \asymuncert{0.117}{0.017}{0.012} & &        &                &         &         &       &           &          \\              
 8.391 & $1^-$ & & \asymuncert{8.5}{0.6}{0.6}&  8.395 & $2^+$  &   &  8.395 & $1^-$ & $1^-$ & $1^-$ & & 4.05(38)   \\
 8.417 &$1^+$\footnotemark[1]&\asymuncert{0.104}{0.014}{0.012} &&        &                &         &         &       &           &          \\
 8.461 &$1^+$\footnotemark[1]& \asymuncert{0.21}{0.03}{0.03} &&  8.475 & $2^-$          &         &  8.461  & $1^+$ & $1^+$ & $1^+$ &  0.382(21) &          \\
 \multirow{2}{*}{ 8.513\footnotemark[5]} &  \multirow{2}{*}{$1^-$}     & \multirow{2}{*}{} &\multirow{2}{*}{\asymuncert{19.5}{1.4}{1.3}} &  \multirow{2}{*}{8.516} & \multirow{2}{*}{$1^+$} & \multirow{2}{*}{1.04(15)}&  8.513 & & & $1^+$ &  0.138(33) \\ 
 & & & &  &   & &  8.514 & $1^-$&$1^-$&$1^-$&   & 1.67(37) \\
 8.602 &$1^+$\footnotemark[1]& \asymuncert{0.34}{0.05}{0.04}&&  8.601 & $1^+$          & 0.44(5) &  8.601 & $1^+$ & $1^+$ & $1^+$ &  0.328(33)   &          \\
 8.678 &$1^+$\footnotemark[1] & \asymuncert{0.73}{0.10}{0.08}& &  8.680 & $1^+$          & 0.47(3) &  8.679 & $1^+$ & $1^+$ & &  0.815(41)   &          \\                 
 8.811 & $1$  & \asymuncert{0.029}{0.004}{0.005}&  &  8.817 & $1^+$, $(2^-)$ & 0.19(2) &         &       &           &          \\
 8.856 &$1^+$\footnotemark[1]&  \asymuncert{0.31}{0.05}{0.04}&&  8.854 & $2^+$, $3^-$   &         &  8.857 & $1$ & $1^+$ &  $1^+$ &   0.281(55)  &          \\
 \hspace{0.75ex}8.881\footnotemark[5] &  $1^{(-)}$ & & \asymuncert{12.6}{1.0}{0.9} &  8.875 & $1^+$          & 0.51(4) &  8.880 & $1^-$ & $1^-$ & $1^-$ &    & 4.79(18) \\
 8.959 &$1^+$\footnotemark[1]&  \asymuncert{0.22}{0.03}{0.03}& & 8.967 & $1^+$, $(2^-)$ & 0.23(6) &  8.961 & $1$ &  $1^+$ & $1^+$ &  0.136(14) &          \\
 8.988 &  $1^+$     & \asymuncert{0.090}{0.012}{0.010}& &        &                &         &         &       &           &          \\  
 9.044 &  $1$     & \asymuncert{0.088}{0.012}{0.010} &  &  9.037 & $1^+$, $(2^-)$ & 0.30(4) &         &       &           &          \\
 9.073 &$1^+$\footnotemark[1]& \asymuncert{0.22}{0.03}{0.00} &&  9.073 & $1^+$          & 0.26(5) &  9.073 & $1^{(+)}$ & $1^+$  & $1^+$ &  0.309(21) &          \\
 9.159 &$1^+$\footnotemark[1]& \asymuncert{0.22}{0.04}{0.03}&&  9.163 & $1^+$          & 0.23(3) &  9.157 & $1^+$& $1^+$ & $1^+$ &  0.201(27) &          \\
 9.191 &    $1^-$   & &\asymuncert{13.0}{0.9}{0.9} &        &                &         &  9.191 & $1^-$ & $1^-$ & $1^-$ &   & 2.92(28) \\
 9.245 &$1^+$\footnotemark[1]& \asymuncert{0.126}{0.017}{0.013} &&        &                &         &         &       &           &          \\
 9.268 & $1$      & \asymuncert{0.034}{0.005}{0.004}& \asymuncert{2.8}{0.3}{0.3} &        &                &         &         &       &           &          \\
 9.292 & $1$ & \asymuncert{0.040}{0.006}{0.005}& \asymuncert{3.3}{0.3}{0.3} &        &                &         &         &       &           &          \\
 9.326 & $1^+$\footnotemark[1]& \asymuncert{0.139}{0.017}{0.015} &&        &                &         &  9.326 & $(1)$& $1^+$ & $1^+$ &  0.312(20) &          \\
 \hspace{0.75ex}9.361\footnotemark[5] &   $1^-$    & & \asymuncert{7.6}{0.5}{0.6} &  9.368 & $1^+$, $(2^-)$ & 0.34(4) &  9.369 & $1^{(-)}$ & $1^{-}$ & $1^{-}$ & &  4.32(40)   \\ 
 9.421 &   $(1)$    &\asymuncert{0.029}{0.004}{0.004} & \asymuncert{2.98}{0.19}{0.23}&  9.407 & $(1^+)$, $2^-$ &         &         &       &           &          \\
 9.463 &   $(1)$    & & &  9.468 &                &         &  9.455 &  $1$ & & &  0.068(13) & 0.75(15) \\
 \hspace{0.75ex}9.529\footnotemark[5] &$1$\footnotemark[1]& \asymuncert{0.27}{0.04}{0.03}& \asymuncert{22.6}{1.6}{1.4}&  9.513 & $1^+$, $(2^-)$ & 0.22(15)&  9.523 & $1^-$ & & $1^-$   & &    7.47(48) \\
 9.554 &  $1^-$     & & \asymuncert{9.1}{0.6}{0.7} &  9.552 & $(2^-)$        &         &  9.554  & $1$ & &  $1^-$&  &  4.48(28) \\
 9.635 &  $1$     & \asymuncert{0.088}{0.012}{0.010}& \asymuncert{7.6}{0.5}{0.6} &  9.643 & $(1^+)$, $2^-$ &         &  9.631 & $1$& & $1^-$ & & 3.82(74)   \\
 9.667 &  $1^-$     & &\asymuncert{7.6}{0.6}{0.5} &  9.667 & $2^-$          &         &  9.668 &  $1$ & & $1^-$&  & 2.57(82)   \\ 
 9.728 &  $1^{(-)}$     & & &  &  &         &  9.723  & $1^{(-)}$ & & $1^-$&       &   5.49(81) \\
 9.746 &  $1$     & \asymuncert{0.176}{0.022}{0.020} & &  9.755 & $1^+$, $(2^-)$ & 0.32(5) &         &       &           &          \\ 
 9.842 &$1^+$\footnotemark[1]& \asymuncert{0.25}{0.04}{0.03}& &  9.846 & $1^+$          & 0.54(7) &         &       &           &          \\
 9.870 & $(1)$      & \asymuncert{0.039}{0.006}{0.005} & \asymuncert{4.3}{0.3}{0.4} &  9.870 & $(2^-)$        &         &         &       &           &          \\
 9.921 &   $(1)$    &\asymuncert{0.028}{0.004}{0.004} & \asymuncert{3.0}{0.3}{0.3}&        &                &         &         &       &           &          \\              
 9.952 & $1$      & \asymuncert{0.057}{0.008}{0.006}& \asymuncert{5.1}{0.4}{0.4}&  9.941 & $(1^-)$, $2^+$ &         &         &       &           &          \\
\hspace{0.75ex}10.043\footnotemark[5] &   $1^-$    & & \asymuncert{7.9}{0.7}{0.6} & 10.036 & $(2^-)$   \\
\hspace{0.75ex}10.071\footnotemark[5] &    $1^{(-)}$   & \asymuncert{0.062}{0.010}{0.007} & \asymuncert{6.0}{0.5}{0.6} & 10.073 &  $1^+$          & 0.35(3)    \\
10.095 &  $1$     &  & &  \multirow{2}{*}{10.105} & \multirow{2}{*}{$1^+$} & \multirow{2}{*}{0.21(2)} &   \\          
10.119 &$1^+$\footnotemark[1]& \asymuncert{0.106}{0.015}{0.012}& & & &  \\  
10.159 &$1^+$\footnotemark[1]& \asymuncert{0.143}{0.020}{0.015}& & 10.157 & $1^+$          & 0.37(4) &         &       &           &          \\   
10.214 &$1^+$\footnotemark[1]&\asymuncert{0.30}{0.04}{0.04} & & 10.218 & $1^+$          & 0.56(4) &         \\
10.240 & $1^{(-)}$ & \asymuncert{0.133}{0.019}{0.015} & \asymuncert{12.5}{0.9}{0.9}&            \\
10.310 &  $1^{(-)}$     & & \asymuncert{8.3}{0.6}{0.6} &        &                &         &      \\
%
%
10.377 &   $(1)$    &  \asymuncert{0.030}{0.005}{0.004}& & 10.385 & $1^+$, $(2^-)$ & 0.15(3) &         &       &           &          \\
10.423 & $1$      & \asymuncert{0.048}{0.007}{0.006}& \asymuncert{4.9}{0.4}{0.4} & \multirow{2}{*}{10.438} & \multirow{2}{*}{$(1^-)$, $2^+$} &         &         &       &           &          \\
10.453 &  $1^{(+)}$     & \asymuncert{0.042}{0.006}{0.005}& \asymuncert{4.0}{0.3}{0.4} &  &  &        &         &       &           &          \\
10.494 & $1^+$\footnotemark[1]&\asymuncert{0.177}{0.024}{0.018} & &        &                &               \\
\hspace{0.75ex}10.515\footnotemark[5] & $1^-$      & & \asymuncert{14.0}{1.0}{1.0} & 10.514 & $1^+$          & 0.40(3) &         \\                  
%
%
10.616 & $(1)$      & \asymuncert{0.023}{0.004}{0.003}& \asymuncert{2.7}{0.3}{0.3} & \multirow{2}{*}{10.633} & \multirow{2}{*}{$1^+$} & \multirow{2}{*}{0.32(12)}&         &       &           &          \\
10.643 & $1^{(-)}$      & \asymuncert{0.162}{0.024}{0.017} &  \asymuncert{16.0}{1.2}{1.0}&  &  & &   \\  
10.668 &$1^+$\footnotemark[1]&\asymuncert{1.13}{0.16}{0.12} &  & 10.670 & $1^+$          & 1.25(6)         \\
10.688 & $1$ & \asymuncert{0.112}{0.015}{0.013} & \asymuncert{11.3}{1.0}{0.9} \\
10.713 & $1^{(-)}$      & \asymuncert{0.058}{0.009}{0.007} & \asymuncert{6.2}{0.6}{0.5} &        &                &         &        \\
10.735 & $1$      & \asymuncert{0.051}{0.008}{0.006} &\asymuncert{5.1}{0.5}{0.5} & 10.735 &                &        \\
%
%
%
%
10.993 &   $(1)$    & & & \multirow{2}{*}{11.013} &    \multirow{2}{*}{$1^+$}       &  \multirow{2}{*}{0.57(3)}&         &       &           &          \\
11.011 &   $1^+$    & \asymuncert{0.182}{0.026}{0.019}& &  & &  &     \\
11.037 &  $1^{(+)}$     & \asymuncert{0.053}{0.008}{0.006}& \asymuncert{5.5}{0.5}{0.4} & 11.041 & $(2^+)$        &         &         &       &           &          \\
11.062 &   $1$    & \asymuncert{0.032}{0.005}{0.004}& \asymuncert{3.4}{0.3}{0.3}& \multirow{2}{*}{11.080} & \multirow{2}{*}{$1^+$, $(2^-)$} & \multirow{2}{*}{0.22(7)} &         &       &           &          \\
11.080 &   $1$    & \asymuncert{0.059}{0.007}{0.008} & \asymuncert{7.0}{0.5}{0.5}& & & &         &       &           &          \\
%
%
11.169 &$1^+$\footnotemark[1]& \asymuncert{0.063}{0.009}{0.007} &11.160  &$2^+$, $3^-$ & &         &         &       &           &          \\ 
11.299 &  $1^{(-)}$     & \asymuncert{0.048}{0.007}{0.006}&  \asymuncert{5.5}{0.4}{0.5}& 11.297 & $2^+$          &         &         &       &           &          \\
11.389 &  $1$     & \asymuncert{0.037}{0.006}{0.004}& \asymuncert{4.0}{0.4}{0.3}&        &                &         &         &       &           &          \\
11.418 &$1^+$\footnotemark[1]& \asymuncert{0.103}{0.014}{0.012} & & 11.410 & $2^+$, $3^(-)$ &         &         &       &           &          \\
11.438 &  $(1)$     & \asymuncert{0.052}{0.008}{0.006}& &        &                &         &         &       &           &          \\             
11.545 &  $(1)$     & \asymuncert{0.028}{0.004}{0.004}& \asymuncert{3.1}{0.3}{0.3}& 11.536 & $(1^+)$, $2^-$ &         &         &       &           &          \\
11.566 & $1$      & \asymuncert{0.076}{0.011}{0.009} & \asymuncert{8.9}{0.7}{0.6} &        &                &         &         \\
%
%
11.632 &  $1$     & \asymuncert{0.051}{0.007}{0.006} & \asymuncert{5.8}{0.4}{0.5} & 11.639 & $2^+{,}3^-$    &         &            &          \\             
11.676 &$1^+$\footnotemark[1]& \asymuncert{0.21}{0.03}{0.03}& & 11.680 & $1^+$          & 0.17(3) &         &       &           &          \\
11.706 &  $1$     & \asymuncert{0.043}{0.006}{0.005}& \asymuncert{4.9}{0.4}{0.4} &        &                &         &         &       &           &          \\
11.806 &   $1^{(-)}$    & &\asymuncert{7.4}{0.6}{0.6} & 11.800 & $(2^+)$        &         &       \\           
%
%
11.856 &  $(1)$     & \asymuncert{0.056}{0.008}{0.007}& & 11.860 & $1^+$          & 0.43(29)&         &       &           &          \\
11.889 &$1^+$\footnotemark[1]& \asymuncert{0.21}{0.03}{0.03}& & 11.890 & $(1^+)$, $2^-$ &         &         &       &           &          \\
%
%
12.047 & $(1)$      &\asymuncert{0.042}{0.006}{0.005} & \asymuncert{5.0}{0.4}{0.5}& 12.040 & $2^+$          &         &         &       &           &          \\
12.074 &  $(1)$     & \asymuncert{0.023}{0.004}{0.004}& & 12.090 &                &         &         &       &           &          \\
12.206 &  $1$     & \asymuncert{0.141}{0.020}{0.016} & \asymuncert{19.3}{1.3}{1.3}& 12.197 & $2^+$          &         &         &       &           &          \\
\multirow{2}{*}{12.268} &   \multirow{2}{*}{$1$}     &\multirow{2}{*}{\asymuncert{0.041}{0.008}{0.005}} & \multirow{2}{*}{\asymuncert{5.6}{0.5}{0.5}} & 12.249 &                &         &         &       &           &          \\
 &      & & & 12.280 &  $1^-$         &         &         &       &           &          \\
12.300 &  $1^{(-)}$     &\asymuncert{0.158}{0.023}{0.017} & \asymuncert{20.7}{1.6}{1.4} &        &                &         &         &       &           &          \\
12.358 &   $(1)$    & \asymuncert{0.038}{0.005}{0.005}& \asymuncert{4.8}{0.4}{0.4}&        &                &         &         &       &           &          \\             
12.392 &  $1$     & \asymuncert{0.080}{0.011}{0.010}& \asymuncert{10.0}{0.9}{0.8}& 12.368 &     $(2^-)$    &         &         &       &           &          \\
12.414 &  $1^{(-)}$     & \asymuncert{0.084}{0.012}{0.010}& \asymuncert{10.5}{0.8}{0.9}&        &                &         &         &       &           &          \\              
12.448 &  $1$     & \asymuncert{0.065}{0.011}{0.009}& \asymuncert{8.2}{1.1}{0.9}& 12.447 &  $(2^+)$       &         &         &       &           &          \\
12.472 & $(1)$      & \asymuncert{0.039}{0.008}{0.007}& \asymuncert{4.9}{0.8}{0.8}& 12.482 &  $(2^+$, $4^+)$&         &         &       &           &          \\
12.574 &   $1^{(-)}$    & &  \asymuncert{9.3}{0.7}{0.8}& 12.573 &  $2^+$, $3^-$  &         &         &       &           &          \\
12.605 &   $1^{(-)}$    & & & 12.613 &       $2^+$    &         &         &       &           &          \\
12.637 &  $1$     & \asymuncert{0.054}{0.008}{0.007}& \asymuncert{7.6}{0.7}{0.7} &  \multirow{2}{*}{12.647} &  \multirow{2}{*}{$2^+$, $(4^+)$} &         &         &       &           &          \\
12.660 &  $1^{(-)}$     & & \asymuncert{14.3}{1.1}{1.0}&  &  &         &         &       &           &          \\
12.750 &$1^+$\footnotemark[1]& \asymuncert{0.240}{0.040}{0.030}& & 12.746 &   $(2^-)$      &         &         &       &           &          \\
12.773 &  $1$     & \asymuncert{0.087}{0.012}{0.010} & \asymuncert{12.0}{0.9}{0.9}&        &                &         &         &       &           &          \\              
12.801 &   $1$    & \asymuncert{0.081}{0.011}{0.009}& & 12.796 & $1^+$, $(2^-)$ & 0.47(9) &         &       &           &          \\
12.819 &   $1$    & \asymuncert{0.058}{0.008}{0.007}& \asymuncert{7.7}{0.7}{0.6} & 12.837 &  $(2^+)$       &         &         &       &           &          \\
12.846 &    $1$  & \asymuncert{0.087}{0.013}{0.010}&  \asymuncert{11.5}{0.9}{1.1}& \multirow{2}{*}{12.858} &  \multirow{2}{*}{$2^+$}      &         &         &       &           &          \\
12.874 &  $1$     & \asymuncert{0.064}{0.010}{0.009}& \asymuncert{8.4}{0.9}{1.0}&  &         &         &         &       &           &          \\
12.892 &   $1$    & \asymuncert{0.030}{0.006}{0.005} & \asymuncert{4.2}{0.7}{0.7} &        &                &         &         &       &           &          \\         
12.966 &   $1^{(-)}$    & & \asymuncert{7.3}{0.6}{0.6} &  \multirow{2}{*}{12.971} &     \multirow{2}{*}{$2^+$}       &         &         &       &           &          \\
12.983 & $(1)$      & \asymuncert{0.038}{0.006}{0.005}& \asymuncert{5.2}{0.5}{0.5} &  &         &         &         &       &           &          \\
13.029 &$1$       & \asymuncert{0.050}{0.007}{0.006} & \asymuncert{6.7}{0.6}{0.5}&  \multirow{2}{*}{13.022} &   \multirow{2}{*}{$2^+$, $4^+$}  &         &         &       &           &          \\
13.034 &  $1$     & \asymuncert{0.062}{0.009}{0.007}& \asymuncert{8.4}{0.7}{0.7}&  &    &         &         &       &           &          \\
13.090 &  $1^{(-)}$     & \asymuncert{0.165}{0.023}{0.018} & \asymuncert{24.2}{1.6}{1.7}&        &                &         &         &       &           &          \\
13.175 &  $1^{(+)}$     & \asymuncert{0.088}{0.012}{0.010}& & 13.176 & $1^+$, $(2^-)$ & 0.37(6) &         &       &           &          \\
 \multirow{2}{*}{13.251} &   \multirow{2}{*}{$1$}     & \multirow{2}{*}{\asymuncert{0.082}{0.012}{0.009}}& \multirow{2}{*}{\asymuncert{12.1}{1.0}{0.9}}& 13.233 &   $2^+$        &         &         &       &           &          \\
 &       & & & 13.260 & $2^+$          &         &         &       &           &          \\ 
13.329 &  $1$      & & & 13.345 &   $2^+$        &         &         &       &           &          
\end{longtable*}
\onecolumngrid
\begin{minipage}{1.5\textwidth}
	\footnotetext[1]{Isobaric analog state in \textsuperscript{58}Cu reported in Ref.~\cite{fuj07}.}
	\footnotetext[2]{Ref.~\cite{bau00}.}
	\footnotetext[3]{Ref.~\cite{sch13}.}
	\footnotetext[4]{Ref.~\cite{shi24}.}
    \footnotetext[5]{Candidates for toroidal $E1$ transitions \cite{vnc23}.} 
\end{minipage}
\ \\
\twocolumngrid
\normalsize
}


The results of the state-by-state analysis are summarized in Table~\ref{tab:all} and compared to high-resolution studies of dipole strength with polarized photon \cite{bau00,sch13,shi24} and electron \cite{met87} scattering.
Excited states observed in the latter experiments are considered being the same as in the present data when they satisfy the condition \cite{wei94}
\begin{equation}
	\frac{|E_{\rm x,1} - E_{\rm x,2}|}{\sqrt{u^2_1(E_{\rm x,1})+ u^2_1(E_{\rm x,2})}} \leq \sqrt{2}.
	\label{eq:root}
\end{equation}
Here, $E_{{\rm x},1(2)}$ denote the excitation energies from the transitions found in experiment 1(2) and $u_{1(2)}(E_{{\rm x},1(2)})$ are the respective uncertainties in excitation energy.

In total 116 dipole transitions were identified based on their forward-peaked angular distribution in the $(p,p^\prime)$ experiment.
26 of them allow an unambiguous $M1$, 11 an $E1$ assignment, and 22 a preference for either multipolarity with the procedure described above. 
The parity of 57 $J=1$ levels could not be unambiguously determined from the present data. 

Parity assignments of the excited dipole states are also possible from the other experiments. 
Polarization of the incident photon beam permits a distinction between $E1$ and $M1$ transitions by the comparison of in-plane and out-of-plane measurements of the scattered photon \cite{zil22}.
Especially laser Compton backscattering-generated photon beams \cite{ama09,wel09} with 100\% linear polarization show unprecedented sensitivity for an unique multipolarity determination. 
The $(e,e^\prime)$ experiment was performed  at low incident energies and backward angles \cite{met87}.
These kinematics favor $M1$ (and to a lesser extent $M2$) transitions which show increasing cross sections with scattering angle, while $E1$ transitions typically show small cross sections at backward angles.

\begin{figure} 
\centering
\includegraphics[width=\columnwidth]{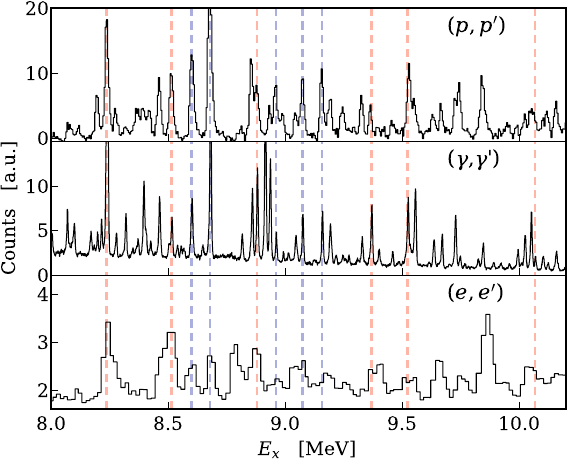}
\caption{Comparison of spectra from the $^{58}$Ni($p,p^\prime$), ($\gamma,\gamma^\prime$) and ($e,e^\prime$) reactions in the excitation energy range $8 -10$ MeV for kinematics where dipole excitations are enhanced.
The vertical lines indicate dipole transitions seen in all three experiments.
Their $E1$ (red) or $M1$ (blue) character is based on the combined analysis of the ($p,p^\prime$) and ($\gamma,\gamma^\prime$) data.
}
\label{fig:spectra-comparison}
\end{figure}

The identification of related transitions according to Eq.~(\ref{eq:root}) in the three reactions is illustrated in Fig.~\ref{fig:spectra-comparison} combining a $(p,p^\prime)$ spectrum at the most forward angle with an $(e,e^\prime)$ spectrum at the most backward angle and unpublished $(\vec{\gamma},\vec{\gamma}^\prime)$ data measured at HI$\gamma$S \cite{wel09}.
For better visibility, the excitation energy region covered is limited to $8 -10$ MeV.
The vertical dashed lines indicate $E1$ (red) or $M1$ (blue) transitions observed in all three reactions.

Finally, IV spinflip-$M1$ transitions in the present data can be identified from comparison to the results of Fujita {\it et al.}~\cite{fuj07}, who studied analog transitions from a $(p,p^\prime)$ experiment at $0^\circ$ and GT transitions populated in the $^{58}$Ni($^3$He,$t$) reaction.
Furthermore, the conversion of cross sections to GT strength contains isospin Clebsch-Gordan coefficients \cite{fuj11}, which allow to determine the final-state isospin from the cross section ratio in the two experiments.  
Cross sections for the $(p,p^\prime)$ reaction at $E_0 = 160$ MeV and extrapolated to $0^\circ$ are given in Table~I of Ref.~\cite{fuj07}. 
The cross section ratio for the population of the same states at a fixed angle in the present experiment should be constant.

The comparison in Table~\ref{tab:all} reveals several cases of unambiguous $E1$ assignments from the present work and the $(\vec{\gamma},\vec{\gamma}^\prime)$ data identified as $M1$ in the the $(e,e^\prime)$ data.
Since $E1$ modes like the IS compressional and IV giant resonances are irrotational, their transverse form factors are weak.
Accordingly, dipole transitions with large transverse cross sections were considered to be of $M1$ nature only in Ref.~\cite{met87}.
However, recent work \cite{vnc23} shows that these states likely belong to a toroidal electric dipole mode \cite{rep19} with unusual properties like large transverse electron scattering form factors mimicking the behavior of $M1$ transitions. 

In general, there is fair agreement between the multipolarity assignments from the different experiments.
However, there is an unclear assignment already noted in Ref.~\cite{fuj07} with respect to the state seen at 9.529 MeV in the present experiment with corresponding states at 9.513 MeV in  $(e,e^\prime)$ and at 9.523 MeV in $(\gamma,\gamma^\prime)$.
The $(\vec{\gamma},\vec{\gamma}^\prime)$ data \cite{bau00,shi24} unambiguously assign a prominent transition with $E1$ character.
On the other hand, the CE data described in Ref.~\cite{fuj07} see a possible analog state which would point towards a $M1$ character.
The transition seen in the $(p,p^\prime)$ experiments (present work and Ref.~\cite{fuj07}) might thus represent an unresolved doublet.
However, the absence of any $M1$ signal in the $(\vec{\gamma},\vec{\gamma}^\prime)$ experiments led us to assume dominance of the $E1$ transition making it part of the toroidal mode discussed above.
Furthermore, according to Ref.~\cite{shi24}, the transition at 8.513 MeV identified as toroidal candidate represents an unresolved $E1$/$M1$ doublet.

\subsection{E1 strength}

\begin{figure}
    \centering
    \includegraphics[width =\columnwidth]{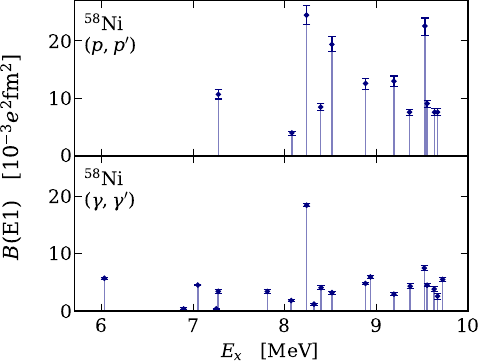}
    \caption{Comparison of $B(E1)$ strengths in $^{58}$Ni deduced from the present work and $(\vec{\gamma},\vec{\gamma}^\prime)$ experiments \cite{bau00,shi24}}.
    \label{fig:be1}
\end{figure}

A comparison of $B(E1)$ strengths deduced from the present experiment up to 10 MeV with results from the $(\vec{\gamma},\vec{\gamma}^\prime)$ experiments \cite{bau00,shi24} is presented in Fig.~\ref{fig:be1}.
Transitions with unique $E1$ character in both experiments or with a $J = 1$ assignment in the former and an unique $E1$ assignment in the latter data are included. 

In general, larger $B(E1)$ strengths are found in the present experiment.
The conversion of $(\gamma,\gamma^\prime)$ cross sections to transition strengths depends on the unknown branching ratio of the g.s.\ decay.
Except for a few cases of direct decay to low-lying states observed and included, the $(\gamma,\gamma^\prime)$ transitions strengths in Table~\ref{tab:all} and Fig.~\ref{fig:be1} assume 100\% g.s.\ decay. 
Further, competition of proton emission is expected above the threshold ($S_p = 8.172$ MeV). 
Thus, the $B(E1)$ values from $(\gamma,\gamma^\prime)$ represent lower limits only.
The average difference to the present results is about a factor of 2. 
The most prominent transition at 8.236 MeV is an exception where the difference is smaller than 25\%.

\subsection{M1 strength}

\begin{figure}
    \centering
    \includegraphics[width =\columnwidth]{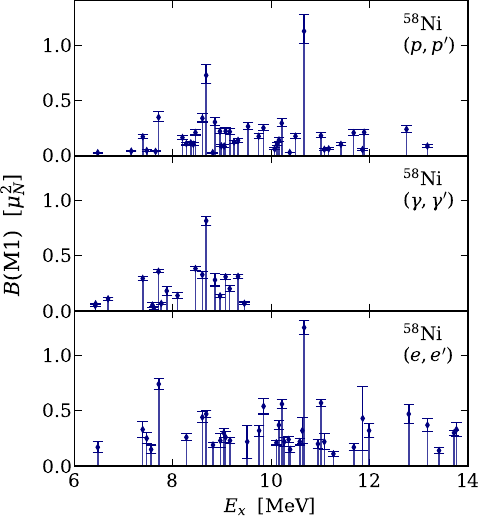}
    \caption{Comparison of $B(M1)$ strengths in $^{58}$Ni deduced from the present work, $(\vec{\gamma},\vec{\gamma}^\prime)$ \cite{bau00,shi24} and $(e,e^\prime)$ \cite{met87} experiments.
    }
    \label{fig:bm1}
\end{figure}

The comparison of the $B(M1)$ strength distributions extracted from the  $(p,p^\prime)$, $(\gamma,\gamma^\prime)$ and $(e,e^\prime)$ experiments is shown in Fig.~\ref{fig:bm1}. 
The $B(M1)$ values from the present work were obtained with the aid of Eq.~(\ref{eq:em_BM1}), i.e., neglecting orbital contributions.
While this may be a reasonable assumption for the total strength because of the random sign of the interference term \cite{fay97}, differences  are expected for individual transitions when comparing with electromagnetic probes.
As discussed for the case of $E1$ transitions above, the strengths deduced from the $(\gamma,\gamma^\prime)$ data are lower limits only.

In the ($\gamma$, $\gamma'$) experiments only the strength up to 9.3~MeV was measured \cite{bau00,shi24}. 
While the $(p,p^\prime)$ experiment is mainly sensitive to the spin part and isovector excitations, the strength distributions from $(e,e^\prime)$ and ($\gamma,\gamma'$) consist of spin and orbital contributions as well as their interference. 
The ($\gamma,\gamma^\prime$) strength distribution shows a good agreement with the present results indicating small orbital strength and dominant g.s.\ back decay. 
A similar behavior was observed in $^{60}$Ni and successfully interpreted in the framework of the quasiparticle-phonon model \cite{sch13}.
The strong transition at 8.678 MeV  is observed with similar strength as in the $(\gamma,\gamma^\prime)$ experiment, while electron scattering gives a value about a factor of 2 smaller. 
On the other hand, the state at 7.715 MeV is enhanced in $(e,e^\prime)$ compared to the other experiments. 
The agreement between the $(p,p^\prime)$  and $(e,e^\prime)$ strength distributions increases with excitation energy, even though many of the transitions observed in the latter are slightly larger.

\subsection{Spinflip-$M1$ strength}

\begin{figure}
\centering
\includegraphics[width=\columnwidth]{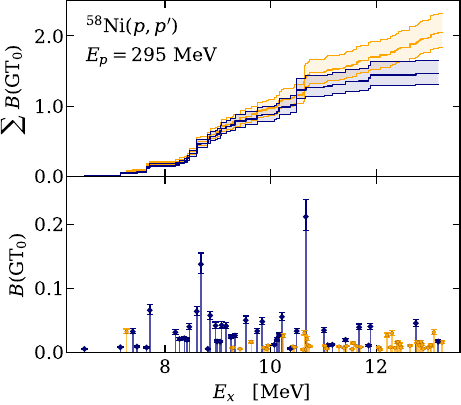}
\caption{Bottom: Distribution of the isovector spin-flip $M1$ transition strength in $^{58}$Ni. 
Blue triangles refer to transitions with uniquely defined $M1$ character, orange squares to tentative assignments.
Top: Running sum excluding (blue) or including (orange) tentative assignments.}
\label{fig:bm1_pp}
\end{figure}

The bottom part of Fig.~\ref{fig:bm1_pp} presents the distribution of IV spinflip-$M1$ strength extracted with the method described in Sec.~\ref{subsec:conversion}.
Blue triangles denote unique $M1$ assignments using the same criteria as discussed for the $E1$ case.
Orange squares are tentative assignments ($J = 1$ in Table~\ref{tab:all}).
The latter are weak and typically below the sensitivity limit of the $(e,e^\prime)$ experiment due to the strong radiative tail background in the spectra of Ref.~\cite{met87}.
The upper part of Fig.~\ref{fig:bm1_pp} shows the running sum excluding (blue) or including (orange) tentative assignments.
Their contribution is negligible up to 9 MeV and reaches about 20\%(40\%) at 12(13) MeV.

\begin{figure}
    \centering
    \includegraphics[width =\columnwidth]{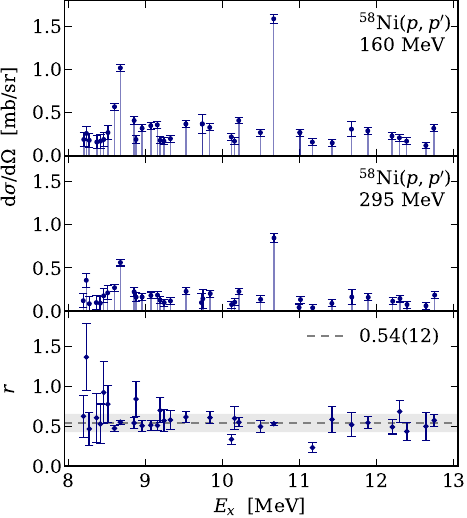}
    \caption{
    Cross sections of dipole transitions observed in the $(p,p^\prime)$ reaction at $\theta_{\rm lab} = 0.4^\circ$ in the present work and in Ref.~\cite{fuj07}. 
    The horizontal dashed line shows the average ratio expected to be constant for $M1$ transitions, cf.\ Eq.~(\ref{eq:crs_ratio}).
    }
    \label{fig:fujita_vgl}
\end{figure}

As discussed in Sec.~\ref{subsec:comparison_general},
IV spinflip-$M1$ transitions can be assigned independently by comparison with GT strength distributions from high-resolution CE reactions \cite{fuj07}.
Here we compare directly with $(p,p^\prime)$ cross section data given in Table II of Ref.~\cite{fuj07} obtained in the solid-angle range $\Theta_{\rm lab} = 0^\circ - 1^\circ$.
For $M1$ transitions, a constant cross section ratio is expected for any of the present spectra.
The data at $0.4^\circ$ are chosen as representative example.
The corresponding integrated cross sections of resolved transition are displayed in the top and middle row of Fig.~\ref{fig:fujita_vgl}, respectively.
The ratio
\begin{equation}
    \label{eq:crs_ratio}
	r = \frac{\frac{\mathrm{d}\sigma}{\mathrm{d}\Omega}(\mathrm{300 MeV}, 0.4^\circ)}{\frac{\mathrm{d}\sigma}{\mathrm{d}\Omega}(\mathrm{160 MeV},0^\circ-1^\circ)}
\end{equation}
is shown in the bottom row of Fig.~\ref{fig:fujita_vgl}.
The average value $r = 0.54$ is indicated as dashed line.

Clearly, most transitions agree with the average $r$ value within error bars.
The states observed in Ref.~\cite{fuj07} at 9.739 MeV and 11.003 MeV  were left out. 
They each correspond to the sum of two close-lying states resolved in the present work due to the slightly better energy resolution. 
For a few transitions between 8 and 9 MeV, values $r \gg 0.54$ are obtained. 
Comparison with Table~\ref{tab:all} reveals that these are $E1$ transitions. 
Their cross section ratios are consistent with the increase of the virtual photon flux in relativistic Coulomb excitation for the two beam energies. 

\begin{figure}[t]
    \centering
    \includegraphics[width = 0.95\columnwidth]{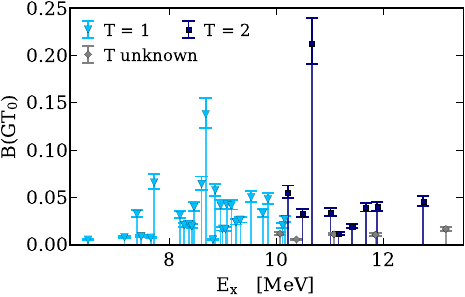}
    \caption{Isospin $T_f$ of final states excited by spin-flip $M1$ transitions in the present work based on the comparison to analog GT transitions studied in Ref.~\cite{fuj07}. 
    The colors indicate $T_f = 1$ (cyan), $T_f = 2$ (blue), not determined (grey).}
    \label{fig:isospins}
\end{figure}

Following the approach of Refs.~\cite{fuj07,fuj11}, one can also assign isospin quantum numbers to the excited $1^+$ states by comparison with the analog GT transitions observed in the high-resolution study of the $^{58}$Ni($^3$He,$t$) reaction presented in Ref.~\cite{fuj07}.
The results are displayed in Fig.~\ref{fig:isospins} and exhibit a two-bump structure of $T_f = 1$ centered around 8.5 MeV and 11 MeV for  $T_f = 1$ and 2 transitions, respectively.
Below an excitation energy in $^{58}$Ni corresponding to the analog of the g.s.\ in $^{58}$Cu only $T_f = 1$ is possible.  
Since all excited states reported in Ref.~\cite{fuj07} between 8 MeV und 13 MeV are also found in the present work, but no additional states  (except the doublets mentioned above), the conclusions regarding the isospin of the excited $1^+$ states remain unchanged with respect to Table~II of Ref.~\cite{fuj07}.  
Some spinflip-$M1$ excitations without an an experimental counterpart in Refs.~\cite{met87,fuj07} were observed in the present data at higher excitation energies, probably due to sensitivity limits of these experiments resulting from the significant background in the spectra.

\section{\label{sec:shell model}Shell model calculations}

\subsection{\label{subsec:shell model details}Details of the calculations}

Calculations within the shell model framework have been performed to assist in the assignment of isospin and interpretation of the $M1$ transitions. 
The model space adopted is composed by the $1f_{7/2}$, $2p_{3/2}$, $1f_{5/2}$ and $2p_{1/2}$ proton and neutron orbitals resting on a $^{40}$Ca core. 
Due to the large dimension of the problem, a truncation was required on the admitted configurations, allowing only configurations with at most 8 particle-hole excitations.
The $fp$-shell effective interactions GXPF1A \cite{hon04,hon05} and KB3G \cite{pov01} have been used. 
The usual quenching factor for Gamow–Teller processes in the $fp$-shell of 0.74 \cite{mar96} has been employed.
The calculations have been performed in the $J$-coupled scheme using a fast implementation of the Lanczos algorithm through the code NATHAN \cite{caurier_shell_2005}.
The results shown correspond to 100 Lanczos iterations.

\subsection{\label{subsec:shell model comparison}Comparison to experiment}

\begin{figure}
    \centering
    \includegraphics[width = \columnwidth]{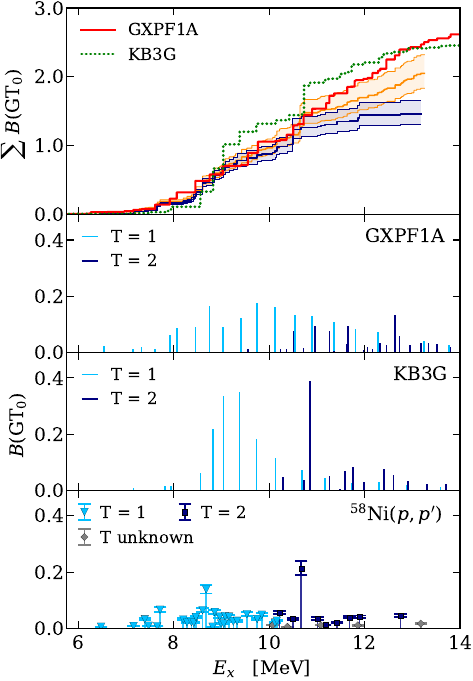}
    \caption{Experimental and theoretical $B(\mathrm{GT_0})$ strength in $^{58}$Ni. 
    Top panel: cumulative sum of the experiment excluding (blue) and including (orange) tentative assignments and shell-model calculations with the GXPF1A (solid red line) and KB3G (green dashed line) interactions. 
    Second and third panel: strength distributions obtained with the GXPF1A \cite{hon04,hon05} and KB3G \cite{pov01} interactions, respectively.  
    Bottom panel: experimental strength distributions from the $(p,p^\prime)$ reaction.
    Final-state isospins are indicated by cyan ($T = 1$) and blue $(T = 2)$.
    }
    \label{fig:gt0_pp}
\end{figure}

Figure \ref{fig:gt0_pp} presents a comparison of the experimental $B(\mathrm{GT_0})$ strength to shell model calculations with both effective interactions. 
Within the limitations of the model approach, which is not sufficient to fully reproduce details of the fragmentation, the KB3G results provide a fair description of the data although the strength in the excitation energy region $8 - 10$ MeV is somewhat overestimated. 
The bimodal shape of the strength distribution due to isospin splitting is reproduced even though the center of mass for $T=1$ and $T=2$ transitions is pushed to slightly higher excitation energies. 
The calculation also reproduces the exceptionally strong $T = 2$ transition at about 10.5 MeV.
The bimodal shape is not clearly present in the GXPF1A result, which produces wider $T = 1$ and $T = 2$ distributions.  
The centroid of the $T=2$ strength is shifted to even higher excitation energies. 

In the top panel of Fig. \ref{fig:gt0_pp}, the cumulative sums are shown. 
Here, the GXPF1A interaction provides a very good description of the data up to 10.5 MeV. 
The KB3G result agrees fairly well up to 9 Mev and overshoots about 20\% up to 10.5 MeV.
At even higher excitation energies up to 13 MeV the model predictions lie above the experimental cumulative sum, even including tentative assignments, cf.~Fig.~\ref{fig:bm1_pp}).
Again, the GXPF1A results are closer to the data.
In this high-energy region it is likely that some strength is missed in the experiment due to the increasing level density.

\begin{figure}
    \centering
    \includegraphics[width = \columnwidth]{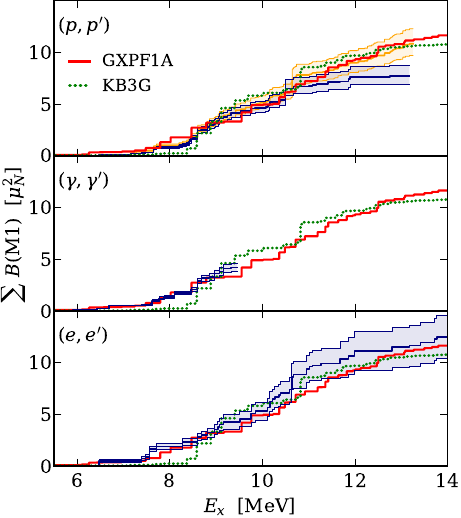}
    \caption{
Running sums of the $B(M1)$ strength in $^{58}$Ni deduced from the present inelastic proton (top), photon \cite{bau00,shi24} (middle), and electron \cite{met87} (bottom) scattering experiments compared to shell-model results with the GXPF1A \cite{hon04,hon05} (red line) and KB3G \cite{pov01} (green line) interactions.
}
    \label{fig:bm1_running_sums}
\end{figure}

Finally, the running sums of the $B(M1)$ strengths from the three experiments are displayed in Fig.~\ref{fig:bm1_running_sums} together with the shell-model results using the GXPF1A and KB3G interactions, respectively.
The differences between the two theoretical results are smaller than for the $B({\rm GT}_0)$ case and they coincide well with all three experimental data sets.
One distinct difference though is the absence of strength in the KB3G result below 8 MeV.
The data do show a small but finite experimental strength amounting to about 0.6 $\mu_N^2$ in the $(p,p^\prime)$  and about 1~$\mu_N^2$ in the $(e,e^\prime)$ results well described by the GXPF1A calculation.

\section{\label{sec:conclusion}Conclusions}

We present here a state-by-state analysis of dipole transitions in $^{58}$Ni measured with the $(p,p^\prime)$ reaction at extreme forward angles including $0^\circ$.
This is possible up to excitation energies of 13 MeV due to the excellent energy resolution of 22 keV (FWHM).
The resulting $E1$ and $M1$ strength distributions are compared to results obtained with the $(e,e^\prime)$ and $(\gamma,\gamma^\prime)$ reactions.

The multipolarity assigments are in fair agreement except for a certain class of transitions unambiguously assigned $E1$ by the proton and photon scattering data but interpreted as $M1$ because of their large transverse cross sections in electron scattering.
They provide the first clear experimental evidence for a toroidal $E1$ mode in nuclei as demonstrated elsewhere \cite{vnc23}.
The $E1$ strengths deduced from the present data are generally larger than those from the $(\gamma,\gamma^\prime)$ experiments indicating sizable inelastic branching ratios. 
This is particularly true above the proton emission threshold.
Effects are less pronounced for $M1$ transitions.

The $B(M1)$ and $B$(GT$_0$) strength distributions are compared to large scale shell-model calculations with the effective GXPF1A and KB3G interactions.
Although details of the strength fragmentation cannot be reproduced because of necessary constraints on the huge model space, cumulative properties like the running sums are well described once a quenching factor 0.74 is included for the spin-isospin part of the $M1$ operator.

\begin{acknowledgments}
We thank the Accelerator Group at RCNP for providing a high-quality dispersion-matched beam  for the experiment. 
This work was supported by the Deutsche Forschungsgemeinschaft (DFG, German Research Foundation) under Contract No.\ SFB 1245 (Project ID No.\ 79384907).
MS acknowledges financial support by the UK-STFC (Grant N0 ST/P005101/1).
\end{acknowledgments}

\bibliography{58Ni-M1}

\begin{thebibliography}{49}%
\makeatletter
\providecommand \@ifxundefined [1]{%
 \@ifx{#1\undefined}
}%
\providecommand \@ifnum [1]{%
 \ifnum #1\expandafter \@firstoftwo
 \else \expandafter \@secondoftwo
 \fi
}%
\providecommand \@ifx [1]{%
 \ifx #1\expandafter \@firstoftwo
 \else \expandafter \@secondoftwo
 \fi
}%
\providecommand \natexlab [1]{#1}%
\providecommand \enquote  [1]{``#1''}%
\providecommand \bibnamefont  [1]{#1}%
\providecommand \bibfnamefont [1]{#1}%
\providecommand \citenamefont [1]{#1}%
\providecommand \href@noop [0]{\@secondoftwo}%
\providecommand \href [0]{\begingroup \@sanitize@url \@href}%
\providecommand \@href[1]{\@@startlink{#1}\@@href}%
\providecommand \@@href[1]{\endgroup#1\@@endlink}%
\providecommand \@sanitize@url [0]{\catcode `\\12\catcode `\$12\catcode
  `\&12\catcode `\#12\catcode `\^12\catcode `\_12\catcode `\%12\relax}%
\providecommand \@@startlink[1]{}%
\providecommand \@@endlink[0]{}%
\providecommand \url  [0]{\begingroup\@sanitize@url \@url }%
\providecommand \@url [1]{\endgroup\@href {#1}{\urlprefix }}%
\providecommand \urlprefix  [0]{URL }%
\providecommand \Eprint [0]{\href }%
\providecommand \doibase [0]{https://doi.org/}%
\providecommand \selectlanguage [0]{\@gobble}%
\providecommand \bibinfo  [0]{\@secondoftwo}%
\providecommand \bibfield  [0]{\@secondoftwo}%
\providecommand \translation [1]{[#1]}%
\providecommand \BibitemOpen [0]{}%
\providecommand \bibitemStop [0]{}%
\providecommand \bibitemNoStop [0]{.\EOS\space}%
\providecommand \EOS [0]{\spacefactor3000\relax}%
\providecommand \BibitemShut  [1]{\csname bibitem#1\endcsname}%
\let\auto@bib@innerbib\@empty
\bibitem [{\citenamefont {Heyde}\ \emph {et~al.}(2010)\citenamefont {Heyde},
  \citenamefont {von Neumann-Cosel},\ and\ \citenamefont {Richter}}]{hey10}%
  \BibitemOpen
  \bibfield  {author} {\bibinfo {author} {\bibfnamefont {K.}~\bibnamefont
  {Heyde}}, \bibinfo {author} {\bibfnamefont {P.}~\bibnamefont {von
  Neumann-Cosel}},\ and\ \bibinfo {author} {\bibfnamefont {A.}~\bibnamefont
  {Richter}},\ }\bibfield  {title} {\bibinfo {title} {Magnetic dipole
  excitations in nuclei: Elementary modes of nucleonic motion},\ }\href
  {https://doi.org/10.1103/RevModPhys.82.2365} {\bibfield  {journal} {\bibinfo
  {journal} {Rev. Mod. Phys.}\ }\textbf {\bibinfo {volume} {82}},\ \bibinfo
  {pages} {2365} (\bibinfo {year} {2010})}\BibitemShut {NoStop}%
\bibitem [{\citenamefont {Fujita}\ \emph {et~al.}(1997)\citenamefont {Fujita},
  \citenamefont {Akimune}, \citenamefont {Daito}, \citenamefont {Fujiwara},
  \citenamefont {Harakeh}, \citenamefont {Inomata}, \citenamefont {J\"anecke},
  \citenamefont {Katori}, \citenamefont {L\"uttge}, \citenamefont {Nakayama},
  \citenamefont {von Neumann-Cosel}, \citenamefont {Richter}, \citenamefont
  {Tamii}, \citenamefont {Tanaka}, \citenamefont {Toyokawa}, \citenamefont
  {Ueno},\ and\ \citenamefont {Yosoi}}]{fuj97}%
  \BibitemOpen
  \bibfield  {author} {\bibinfo {author} {\bibfnamefont {Y.}~\bibnamefont
  {Fujita}}, \bibinfo {author} {\bibfnamefont {H.}~\bibnamefont {Akimune}},
  \bibinfo {author} {\bibfnamefont {I.}~\bibnamefont {Daito}}, \bibinfo
  {author} {\bibfnamefont {M.}~\bibnamefont {Fujiwara}}, \bibinfo {author}
  {\bibfnamefont {M.~N.}\ \bibnamefont {Harakeh}}, \bibinfo {author}
  {\bibfnamefont {T.}~\bibnamefont {Inomata}}, \bibinfo {author} {\bibfnamefont
  {J.}~\bibnamefont {J\"anecke}}, \bibinfo {author} {\bibfnamefont
  {K.}~\bibnamefont {Katori}}, \bibinfo {author} {\bibfnamefont
  {C.}~\bibnamefont {L\"uttge}}, \bibinfo {author} {\bibfnamefont
  {S.}~\bibnamefont {Nakayama}}, \bibinfo {author} {\bibfnamefont
  {P.}~\bibnamefont {von Neumann-Cosel}}, \bibinfo {author} {\bibfnamefont
  {A.}~\bibnamefont {Richter}}, \bibinfo {author} {\bibfnamefont
  {A.}~\bibnamefont {Tamii}}, \bibinfo {author} {\bibfnamefont
  {M.}~\bibnamefont {Tanaka}}, \bibinfo {author} {\bibfnamefont
  {H.}~\bibnamefont {Toyokawa}}, \bibinfo {author} {\bibfnamefont
  {H.}~\bibnamefont {Ueno}},\ and\ \bibinfo {author} {\bibfnamefont
  {M.}~\bibnamefont {Yosoi}},\ }\bibfield  {title} {\bibinfo {title} {Isospin
  and spin-orbital structures of ${J}^{\ensuremath{\pi}}{=1}^{+}$ states
  excited in ${}^{28}${S}i},\ }\href {https://doi.org/10.1103/PhysRevC.55.1137}
  {\bibfield  {journal} {\bibinfo  {journal} {Phys. Rev. C}\ }\textbf {\bibinfo
  {volume} {55}},\ \bibinfo {pages} {1137} (\bibinfo {year}
  {1997})}\BibitemShut {NoStop}%
\bibitem [{\citenamefont {Hofmann}\ \emph {et~al.}(2002)\citenamefont
  {Hofmann}, \citenamefont {von Neumann-Cosel}, \citenamefont {Neumeyer},
  \citenamefont {Rangacharyulu}, \citenamefont {Reitz}, \citenamefont
  {Richter}, \citenamefont {Schrieder}, \citenamefont {Sober}, \citenamefont
  {Fagg},\ and\ \citenamefont {Brown}}]{hof02}%
  \BibitemOpen
  \bibfield  {author} {\bibinfo {author} {\bibfnamefont {F.}~\bibnamefont
  {Hofmann}}, \bibinfo {author} {\bibfnamefont {P.}~\bibnamefont {von
  Neumann-Cosel}}, \bibinfo {author} {\bibfnamefont {F.}~\bibnamefont
  {Neumeyer}}, \bibinfo {author} {\bibfnamefont {C.}~\bibnamefont
  {Rangacharyulu}}, \bibinfo {author} {\bibfnamefont {B.}~\bibnamefont
  {Reitz}}, \bibinfo {author} {\bibfnamefont {A.}~\bibnamefont {Richter}},
  \bibinfo {author} {\bibfnamefont {G.}~\bibnamefont {Schrieder}}, \bibinfo
  {author} {\bibfnamefont {D.~I.}\ \bibnamefont {Sober}}, \bibinfo {author}
  {\bibfnamefont {L.~W.}\ \bibnamefont {Fagg}},\ and\ \bibinfo {author}
  {\bibfnamefont {B.~A.}\ \bibnamefont {Brown}},\ }\bibfield  {title} {\bibinfo
  {title} {Magnetic dipole transitions in ${}^{32}\mathrm{S}$ from electron
  scattering at 180\ifmmode^\circ\else\textdegree\fi{}},\ }\href
  {https://doi.org/10.1103/PhysRevC.65.024311} {\bibfield  {journal} {\bibinfo
  {journal} {Phys. Rev. C}\ }\textbf {\bibinfo {volume} {65}},\ \bibinfo
  {pages} {024311} (\bibinfo {year} {2002})}\BibitemShut {NoStop}%
\bibitem [{\citenamefont {Langanke}\ \emph {et~al.}(2004)\citenamefont
  {Langanke}, \citenamefont {Mart\'{\i}nez-Pinedo}, \citenamefont {von
  Neumann-Cosel},\ and\ \citenamefont {Richter}}]{lan04}%
  \BibitemOpen
  \bibfield  {author} {\bibinfo {author} {\bibfnamefont {K.}~\bibnamefont
  {Langanke}}, \bibinfo {author} {\bibfnamefont {G.}~\bibnamefont
  {Mart\'{\i}nez-Pinedo}}, \bibinfo {author} {\bibfnamefont {P.}~\bibnamefont
  {von Neumann-Cosel}},\ and\ \bibinfo {author} {\bibfnamefont
  {A.}~\bibnamefont {Richter}},\ }\bibfield  {title} {\bibinfo {title}
  {Supernova inelastic neutrino-nucleus cross sections from high-resolution
  electron scattering experiments and shell-model calculations},\ }\href
  {https://doi.org/10.1103/PhysRevLett.93.202501} {\bibfield  {journal}
  {\bibinfo  {journal} {Phys. Rev. Lett.}\ }\textbf {\bibinfo {volume} {93}},\
  \bibinfo {pages} {202501} (\bibinfo {year} {2004})}\BibitemShut {NoStop}%
\bibitem [{\citenamefont {Otsuka}\ \emph {et~al.}(2020)\citenamefont {Otsuka},
  \citenamefont {Gade}, \citenamefont {Sorlin}, \citenamefont {Suzuki},\ and\
  \citenamefont {Utsuno}}]{ots20}%
  \BibitemOpen
  \bibfield  {author} {\bibinfo {author} {\bibfnamefont {T.}~\bibnamefont
  {Otsuka}}, \bibinfo {author} {\bibfnamefont {A.}~\bibnamefont {Gade}},
  \bibinfo {author} {\bibfnamefont {O.}~\bibnamefont {Sorlin}}, \bibinfo
  {author} {\bibfnamefont {T.}~\bibnamefont {Suzuki}},\ and\ \bibinfo {author}
  {\bibfnamefont {Y.}~\bibnamefont {Utsuno}},\ }\bibfield  {title} {\bibinfo
  {title} {Evolution of shell structure in exotic nuclei},\ }\href
  {https://doi.org/10.1103/RevModPhys.92.015002} {\bibfield  {journal}
  {\bibinfo  {journal} {Rev. Mod. Phys.}\ }\textbf {\bibinfo {volume} {92}},\
  \bibinfo {pages} {015002} (\bibinfo {year} {2020})}\BibitemShut {NoStop}%
\bibitem [{\citenamefont {Langanke}\ \emph {et~al.}(2008)\citenamefont
  {Langanke}, \citenamefont {Mart\'{\i}nez-Pinedo}, \citenamefont {M\"uller},
  \citenamefont {Janka}, \citenamefont {Marek}, \citenamefont {Hix},
  \citenamefont {Juodagalvis},\ and\ \citenamefont {Sampaio}}]{lan10}%
  \BibitemOpen
  \bibfield  {author} {\bibinfo {author} {\bibfnamefont {K.}~\bibnamefont
  {Langanke}}, \bibinfo {author} {\bibfnamefont {G.}~\bibnamefont
  {Mart\'{\i}nez-Pinedo}}, \bibinfo {author} {\bibfnamefont {B.}~\bibnamefont
  {M\"uller}}, \bibinfo {author} {\bibfnamefont {H.-T.}\ \bibnamefont {Janka}},
  \bibinfo {author} {\bibfnamefont {A.}~\bibnamefont {Marek}}, \bibinfo
  {author} {\bibfnamefont {W.~R.}\ \bibnamefont {Hix}}, \bibinfo {author}
  {\bibfnamefont {A.}~\bibnamefont {Juodagalvis}},\ and\ \bibinfo {author}
  {\bibfnamefont {J.~M.}\ \bibnamefont {Sampaio}},\ }\bibfield  {title}
  {\bibinfo {title} {Effects of inelastic neutrino-nucleus scattering on
  supernova dynamics and radiated neutrino spectra},\ }\href
  {https://doi.org/10.1103/PhysRevLett.100.011101} {\bibfield  {journal}
  {\bibinfo  {journal} {Phys. Rev. Lett.}\ }\textbf {\bibinfo {volume} {100}},\
  \bibinfo {pages} {011101} (\bibinfo {year} {2008})}\BibitemShut {NoStop}%
\bibitem [{\citenamefont {Gayer}\ \emph {et~al.}(2019)\citenamefont {Gayer},
  \citenamefont {Beck}, \citenamefont {Bhike}, \citenamefont {Isaak},
  \citenamefont {Pietralla}, \citenamefont {Ries}, \citenamefont {Savran},
  \citenamefont {Schilling}, \citenamefont {Tornow},\ and\ \citenamefont
  {Werner}}]{gay19}%
  \BibitemOpen
  \bibfield  {author} {\bibinfo {author} {\bibfnamefont {U.}~\bibnamefont
  {Gayer}}, \bibinfo {author} {\bibfnamefont {T.}~\bibnamefont {Beck}},
  \bibinfo {author} {\bibfnamefont {M.}~\bibnamefont {Bhike}}, \bibinfo
  {author} {\bibfnamefont {J.}~\bibnamefont {Isaak}}, \bibinfo {author}
  {\bibfnamefont {N.}~\bibnamefont {Pietralla}}, \bibinfo {author}
  {\bibfnamefont {P.~C.}\ \bibnamefont {Ries}}, \bibinfo {author}
  {\bibfnamefont {D.}~\bibnamefont {Savran}}, \bibinfo {author} {\bibfnamefont
  {M.}~\bibnamefont {Schilling}}, \bibinfo {author} {\bibfnamefont
  {W.}~\bibnamefont {Tornow}},\ and\ \bibinfo {author} {\bibfnamefont
  {V.}~\bibnamefont {Werner}},\ }\bibfield  {title} {\bibinfo {title}
  {Experimental ${M}1$ response of $^{40}\mathrm{Ar}$ as a benchmark for
  neutrino-nucleus scattering calculations},\ }\href
  {https://doi.org/10.1103/PhysRevC.100.034305} {\bibfield  {journal} {\bibinfo
   {journal} {Phys. Rev. C}\ }\textbf {\bibinfo {volume} {100}},\ \bibinfo
  {pages} {034305} (\bibinfo {year} {2019})}\BibitemShut {NoStop}%
\bibitem [{\citenamefont {Tornow}\ \emph {et~al.}(2022)\citenamefont {Tornow},
  \citenamefont {Tonchev}, \citenamefont {Finch}, \citenamefont {Krishichayan},
  \citenamefont {Wang}, \citenamefont {Hayes}, \citenamefont {Yeomans},\ and\
  \citenamefont {Newmark}}]{tor22}%
  \BibitemOpen
  \bibfield  {author} {\bibinfo {author} {\bibfnamefont {W.}~\bibnamefont
  {Tornow}}, \bibinfo {author} {\bibfnamefont {A.}~\bibnamefont {Tonchev}},
  \bibinfo {author} {\bibfnamefont {S.}~\bibnamefont {Finch}}, \bibinfo
  {author} {\bibnamefont {Krishichayan}}, \bibinfo {author} {\bibfnamefont
  {X.}~\bibnamefont {Wang}}, \bibinfo {author} {\bibfnamefont {A.}~\bibnamefont
  {Hayes}}, \bibinfo {author} {\bibfnamefont {H.}~\bibnamefont {Yeomans}},\
  and\ \bibinfo {author} {\bibfnamefont {D.}~\bibnamefont {Newmark}},\
  }\bibfield  {title} {\bibinfo {title} {Neutral-current neutrino cross section
  and expected supernova signals for $^{40}\mathrm{Ar}$ from a three-fold
  increase in the magnetic dipole strength},\ }\href
  {https://doi.org/https://doi.org/10.1016/j.physletb.2022.137576} {\bibfield
  {journal} {\bibinfo  {journal} {Phys. Lett. B}\ }\textbf {\bibinfo {volume}
  {835}},\ \bibinfo {pages} {137576} (\bibinfo {year} {2022})}\BibitemShut
  {NoStop}%
\bibitem [{\citenamefont {Fujita}\ \emph {et~al.}(2011)\citenamefont {Fujita},
  \citenamefont {Rubio},\ and\ \citenamefont {Gelletly}}]{fuj11}%
  \BibitemOpen
  \bibfield  {author} {\bibinfo {author} {\bibfnamefont {Y.}~\bibnamefont
  {Fujita}}, \bibinfo {author} {\bibfnamefont {B.}~\bibnamefont {Rubio}},\ and\
  \bibinfo {author} {\bibfnamefont {W.}~\bibnamefont {Gelletly}},\ }\bibfield
  {title} {\bibinfo {title} {Spin–isospin excitations probed by strong, weak
  and electro-magnetic interactions},\ }\href
  {https://doi.org/https://doi.org/10.1016/j.ppnp.2011.01.056} {\bibfield
  {journal} {\bibinfo  {journal} {Prog. Part. Nucl. Phys.}\ }\textbf {\bibinfo
  {volume} {66}},\ \bibinfo {pages} {549} (\bibinfo {year} {2011})}\BibitemShut
  {NoStop}%
\bibitem [{\citenamefont {Mart\'{\i}nez-Pinedo}\ \emph
  {et~al.}(1996)\citenamefont {Mart\'{\i}nez-Pinedo}, \citenamefont {Poves},
  \citenamefont {Caurier},\ and\ \citenamefont {Zuker}}]{mar96}%
  \BibitemOpen
  \bibfield  {author} {\bibinfo {author} {\bibfnamefont {G.}~\bibnamefont
  {Mart\'{\i}nez-Pinedo}}, \bibinfo {author} {\bibfnamefont {A.}~\bibnamefont
  {Poves}}, \bibinfo {author} {\bibfnamefont {E.}~\bibnamefont {Caurier}},\
  and\ \bibinfo {author} {\bibfnamefont {A.~P.}\ \bibnamefont {Zuker}},\
  }\bibfield  {title} {\bibinfo {title} {Effective ${g}_{A}$ in the $pf$
  shell},\ }\href {https://doi.org/10.1103/PhysRevC.53.R2602} {\bibfield
  {journal} {\bibinfo  {journal} {Phys. Rev. C}\ }\textbf {\bibinfo {volume}
  {53}},\ \bibinfo {pages} {R2602} (\bibinfo {year} {1996})}\BibitemShut
  {NoStop}%
\bibitem [{\citenamefont {{von Neumann-Cosel}}\ \emph
  {et~al.}(1998)\citenamefont {{von Neumann-Cosel}}, \citenamefont {Poves},
  \citenamefont {Retamosa},\ and\ \citenamefont {Richter}}]{vnc98}%
  \BibitemOpen
  \bibfield  {author} {\bibinfo {author} {\bibfnamefont {P.}~\bibnamefont {{von
  Neumann-Cosel}}}, \bibinfo {author} {\bibfnamefont {A.}~\bibnamefont
  {Poves}}, \bibinfo {author} {\bibfnamefont {J.}~\bibnamefont {Retamosa}},\
  and\ \bibinfo {author} {\bibfnamefont {A.}~\bibnamefont {Richter}},\
  }\bibfield  {title} {\bibinfo {title} {Magnetic dipole response in nuclei at
  the ${N}=28$ shell closure: {A} new look},\ }\href
  {https://doi.org/https://doi.org/10.1016/S0370-2693(98)01298-2} {\bibfield
  {journal} {\bibinfo  {journal} {Phys. Lett. B}\ }\textbf {\bibinfo {volume}
  {443}},\ \bibinfo {pages} {1} (\bibinfo {year} {1998})}\BibitemShut {NoStop}%
\bibitem [{\citenamefont {Gysbers}\ \emph {et~al.}(2019)\citenamefont
  {Gysbers}, \citenamefont {Hagen}, \citenamefont {Holt}, \citenamefont
  {Jansen}, \citenamefont {Morris}, \citenamefont {Navr{\'a}til}, \citenamefont
  {Papenbrock}, \citenamefont {Quaglioni}, \citenamefont {Schwenk},
  \citenamefont {Stroberg},\ and\ \citenamefont {Wendt}}]{gys19}%
  \BibitemOpen
  \bibfield  {author} {\bibinfo {author} {\bibfnamefont {P.}~\bibnamefont
  {Gysbers}}, \bibinfo {author} {\bibfnamefont {G.}~\bibnamefont {Hagen}},
  \bibinfo {author} {\bibfnamefont {J.~D.}\ \bibnamefont {Holt}}, \bibinfo
  {author} {\bibfnamefont {G.~R.}\ \bibnamefont {Jansen}}, \bibinfo {author}
  {\bibfnamefont {T.~D.}\ \bibnamefont {Morris}}, \bibinfo {author}
  {\bibfnamefont {P.}~\bibnamefont {Navr{\'a}til}}, \bibinfo {author}
  {\bibfnamefont {T.}~\bibnamefont {Papenbrock}}, \bibinfo {author}
  {\bibfnamefont {S.}~\bibnamefont {Quaglioni}}, \bibinfo {author}
  {\bibfnamefont {A.}~\bibnamefont {Schwenk}}, \bibinfo {author} {\bibfnamefont
  {S.~R.}\ \bibnamefont {Stroberg}},\ and\ \bibinfo {author} {\bibfnamefont
  {K.~A.}\ \bibnamefont {Wendt}},\ }\bibfield  {title} {\bibinfo {title}
  {Discrepancy between experimental and theoretical $\beta$-decay rates
  resolved from first principles},\ }\href
  {https://doi.org/10.1038/s41567-019-0450-7} {\bibfield  {journal} {\bibinfo
  {journal} {Nat. Phys.}\ }\textbf {\bibinfo {volume} {15}},\ \bibinfo {pages}
  {428} (\bibinfo {year} {2019})}\BibitemShut {NoStop}%
\bibitem [{\citenamefont {Coraggio}\ \emph {et~al.}(2024)\citenamefont
  {Coraggio}, \citenamefont {Itaco}, \citenamefont {De~Gregorio}, \citenamefont
  {Gargano}, \citenamefont {Cheng}, \citenamefont {Ma}, \citenamefont {Xu},\
  and\ \citenamefont {Viviani}}]{cor24}%
  \BibitemOpen
  \bibfield  {author} {\bibinfo {author} {\bibfnamefont {L.}~\bibnamefont
  {Coraggio}}, \bibinfo {author} {\bibfnamefont {N.}~\bibnamefont {Itaco}},
  \bibinfo {author} {\bibfnamefont {G.}~\bibnamefont {De~Gregorio}}, \bibinfo
  {author} {\bibfnamefont {A.}~\bibnamefont {Gargano}}, \bibinfo {author}
  {\bibfnamefont {Z.~H.}\ \bibnamefont {Cheng}}, \bibinfo {author}
  {\bibfnamefont {Y.~Z.}\ \bibnamefont {Ma}}, \bibinfo {author} {\bibfnamefont
  {F.~R.}\ \bibnamefont {Xu}},\ and\ \bibinfo {author} {\bibfnamefont
  {M.}~\bibnamefont {Viviani}},\ }\bibfield  {title} {\bibinfo {title} {The
  renormalization of the shell-model {G}amow-{T}eller operator starting from
  effective field theory for nuclear systems},\ }\href
  {https://doi.org/10.1103/PhysRevC.109.014301} {\bibfield  {journal} {\bibinfo
   {journal} {Phys. Rev. C}\ }\textbf {\bibinfo {volume} {109}},\ \bibinfo
  {pages} {014301} (\bibinfo {year} {2024})}\BibitemShut {NoStop}%
\bibitem [{\citenamefont {Seutin}\ \emph {et~al.}(2023)\citenamefont {Seutin},
  \citenamefont {Hernandez}, \citenamefont {Miyagi}, \citenamefont {Bacca},
  \citenamefont {Hebeler}, \citenamefont {K\"onig},\ and\ \citenamefont
  {Schwenk}}]{seu23}%
  \BibitemOpen
  \bibfield  {author} {\bibinfo {author} {\bibfnamefont {R.}~\bibnamefont
  {Seutin}}, \bibinfo {author} {\bibfnamefont {O.~J.}\ \bibnamefont
  {Hernandez}}, \bibinfo {author} {\bibfnamefont {T.}~\bibnamefont {Miyagi}},
  \bibinfo {author} {\bibfnamefont {S.}~\bibnamefont {Bacca}}, \bibinfo
  {author} {\bibfnamefont {K.}~\bibnamefont {Hebeler}}, \bibinfo {author}
  {\bibfnamefont {S.}~\bibnamefont {K\"onig}},\ and\ \bibinfo {author}
  {\bibfnamefont {A.}~\bibnamefont {Schwenk}},\ }\bibfield  {title} {\bibinfo
  {title} {Magnetic dipole operator from chiral effective field theory for
  many-body expansion methods},\ }\href
  {https://doi.org/10.1103/PhysRevC.108.054005} {\bibfield  {journal} {\bibinfo
   {journal} {Phys. Rev. C}\ }\textbf {\bibinfo {volume} {108}},\ \bibinfo
  {pages} {054005} (\bibinfo {year} {2023})}\BibitemShut {NoStop}%
\bibitem [{\citenamefont {Bauwens}\ \emph {et~al.}(2000)\citenamefont
  {Bauwens}, \citenamefont {Bryssinck}, \citenamefont {De~Frenne},
  \citenamefont {Govaert}, \citenamefont {Govor}, \citenamefont {Hagemann},
  \citenamefont {Heyse}, \citenamefont {Jacobs}, \citenamefont {Mondelaers},\
  and\ \citenamefont {Ponomarev}}]{bau00}%
  \BibitemOpen
  \bibfield  {author} {\bibinfo {author} {\bibfnamefont {F.}~\bibnamefont
  {Bauwens}}, \bibinfo {author} {\bibfnamefont {J.}~\bibnamefont {Bryssinck}},
  \bibinfo {author} {\bibfnamefont {D.}~\bibnamefont {De~Frenne}}, \bibinfo
  {author} {\bibfnamefont {K.}~\bibnamefont {Govaert}}, \bibinfo {author}
  {\bibfnamefont {L.}~\bibnamefont {Govor}}, \bibinfo {author} {\bibfnamefont
  {M.}~\bibnamefont {Hagemann}}, \bibinfo {author} {\bibfnamefont
  {J.}~\bibnamefont {Heyse}}, \bibinfo {author} {\bibfnamefont
  {E.}~\bibnamefont {Jacobs}}, \bibinfo {author} {\bibfnamefont
  {W.}~\bibnamefont {Mondelaers}},\ and\ \bibinfo {author} {\bibfnamefont
  {V.~Y.}\ \bibnamefont {Ponomarev}},\ }\bibfield  {title} {\bibinfo {title}
  {Dipole transitions to bound states in ${}^{56}\mathrm{Fe}$ and
  ${}^{58}\mathrm{Ni}$},\ }\href {https://doi.org/10.1103/PhysRevC.62.024302}
  {\bibfield  {journal} {\bibinfo  {journal} {Phys. Rev. C}\ }\textbf {\bibinfo
  {volume} {62}},\ \bibinfo {pages} {024302} (\bibinfo {year}
  {2000})}\BibitemShut {NoStop}%
\bibitem [{\citenamefont {Scheck}\ \emph {et~al.}(2013)\citenamefont {Scheck},
  \citenamefont {Ponomarev}, \citenamefont {Fritzsche}, \citenamefont
  {Joubert}, \citenamefont {Aumann}, \citenamefont {Beller}, \citenamefont
  {Isaak}, \citenamefont {Kelley}, \citenamefont {Kwan}, \citenamefont
  {Pietralla}, \citenamefont {Raut}, \citenamefont {Romig}, \citenamefont
  {Rusev}, \citenamefont {Savran}, \citenamefont {Schorrenberger},
  \citenamefont {Sonnabend}, \citenamefont {Tonchev}, \citenamefont {Tornow},
  \citenamefont {Weller}, \citenamefont {Zilges},\ and\ \citenamefont
  {Zweidinger}}]{sch13}%
  \BibitemOpen
  \bibfield  {author} {\bibinfo {author} {\bibfnamefont {M.}~\bibnamefont
  {Scheck}}, \bibinfo {author} {\bibfnamefont {V.~Y.}\ \bibnamefont
  {Ponomarev}}, \bibinfo {author} {\bibfnamefont {M.}~\bibnamefont
  {Fritzsche}}, \bibinfo {author} {\bibfnamefont {J.}~\bibnamefont {Joubert}},
  \bibinfo {author} {\bibfnamefont {T.}~\bibnamefont {Aumann}}, \bibinfo
  {author} {\bibfnamefont {J.}~\bibnamefont {Beller}}, \bibinfo {author}
  {\bibfnamefont {J.}~\bibnamefont {Isaak}}, \bibinfo {author} {\bibfnamefont
  {J.~H.}\ \bibnamefont {Kelley}}, \bibinfo {author} {\bibfnamefont
  {E.}~\bibnamefont {Kwan}}, \bibinfo {author} {\bibfnamefont {N.}~\bibnamefont
  {Pietralla}}, \bibinfo {author} {\bibfnamefont {R.}~\bibnamefont {Raut}},
  \bibinfo {author} {\bibfnamefont {C.}~\bibnamefont {Romig}}, \bibinfo
  {author} {\bibfnamefont {G.}~\bibnamefont {Rusev}}, \bibinfo {author}
  {\bibfnamefont {D.}~\bibnamefont {Savran}}, \bibinfo {author} {\bibfnamefont
  {L.}~\bibnamefont {Schorrenberger}}, \bibinfo {author} {\bibfnamefont
  {K.}~\bibnamefont {Sonnabend}}, \bibinfo {author} {\bibfnamefont {A.~P.}\
  \bibnamefont {Tonchev}}, \bibinfo {author} {\bibfnamefont {W.}~\bibnamefont
  {Tornow}}, \bibinfo {author} {\bibfnamefont {H.~R.}\ \bibnamefont {Weller}},
  \bibinfo {author} {\bibfnamefont {A.}~\bibnamefont {Zilges}},\ and\ \bibinfo
  {author} {\bibfnamefont {M.}~\bibnamefont {Zweidinger}},\ }\bibfield  {title}
  {\bibinfo {title} {Photoresponse of ${}^{60}${N}i below 10-{M}e{V} excitation
  energy: Evolution of dipole resonances in $fp$-shell nuclei near ${N}={Z}$},\
  }\href {https://doi.org/10.1103/PhysRevC.88.044304} {\bibfield  {journal}
  {\bibinfo  {journal} {Phys. Rev. C}\ }\textbf {\bibinfo {volume} {88}},\
  \bibinfo {pages} {044304} (\bibinfo {year} {2013})}\BibitemShut {NoStop}%
\bibitem [{\citenamefont {Shizuma}\ \emph {et~al.}(2024)\citenamefont
  {Shizuma}, \citenamefont {Omer}, \citenamefont {Hayakawa}, \citenamefont
  {Minato}, \citenamefont {Matsuba}, \citenamefont {Miyamoto}, \citenamefont
  {Shimizu},\ and\ \citenamefont {Utsuno}}]{shi24}%
  \BibitemOpen
  \bibfield  {author} {\bibinfo {author} {\bibfnamefont {T.}~\bibnamefont
  {Shizuma}}, \bibinfo {author} {\bibfnamefont {M.}~\bibnamefont {Omer}},
  \bibinfo {author} {\bibfnamefont {T.}~\bibnamefont {Hayakawa}}, \bibinfo
  {author} {\bibfnamefont {F.}~\bibnamefont {Minato}}, \bibinfo {author}
  {\bibfnamefont {S.}~\bibnamefont {Matsuba}}, \bibinfo {author} {\bibfnamefont
  {S.}~\bibnamefont {Miyamoto}}, \bibinfo {author} {\bibfnamefont
  {N.}~\bibnamefont {Shimizu}},\ and\ \bibinfo {author} {\bibfnamefont
  {Y.}~\bibnamefont {Utsuno}},\ }\bibfield  {title} {\bibinfo {title} {Parity
  assignment for low-lying dipole states in $^{58}\mathrm{Ni}$},\ }\href
  {https://doi.org/10.1103/PhysRevC.109.014302} {\bibfield  {journal} {\bibinfo
   {journal} {Phys. Rev. C}\ }\textbf {\bibinfo {volume} {109}},\ \bibinfo
  {pages} {014302} (\bibinfo {year} {2024})}\BibitemShut {NoStop}%
\bibitem [{\citenamefont {Mettner}\ \emph {et~al.}(1987)\citenamefont
  {Mettner}, \citenamefont {Richter}, \citenamefont {Stock}, \citenamefont
  {Metsch},\ and\ \citenamefont {{Van Hees}}}]{met87}%
  \BibitemOpen
  \bibfield  {author} {\bibinfo {author} {\bibfnamefont {W.}~\bibnamefont
  {Mettner}}, \bibinfo {author} {\bibfnamefont {A.}~\bibnamefont {Richter}},
  \bibinfo {author} {\bibfnamefont {W.}~\bibnamefont {Stock}}, \bibinfo
  {author} {\bibfnamefont {B.}~\bibnamefont {Metsch}},\ and\ \bibinfo {author}
  {\bibfnamefont {A.}~\bibnamefont {{Van Hees}}},\ }\bibfield  {title}
  {\bibinfo {title} {Electroexcitation of $^{58}\mathrm{Ni}$: A study of the
  fragmentation of the magnetic dipole strength},\ }\href
  {https://doi.org/https://doi.org/10.1016/0375-9474(87)90159-X} {\bibfield
  {journal} {\bibinfo  {journal} {Nucl. Phys. A}\ }\textbf {\bibinfo {volume}
  {473}},\ \bibinfo {pages} {160} (\bibinfo {year} {1987})}\BibitemShut
  {NoStop}%
\bibitem [{\citenamefont {Fujita}\ \emph {et~al.}(2007)\citenamefont {Fujita},
  \citenamefont {Fujita}, \citenamefont {Adachi}, \citenamefont {Bacher},
  \citenamefont {Berg}, \citenamefont {Black}, \citenamefont {Caurier},
  \citenamefont {Foster}, \citenamefont {Fujimura}, \citenamefont {Hara},
  \citenamefont {Harada}, \citenamefont {Hatanaka}, \citenamefont
  {J\"{a}necke}, \citenamefont {Kamiya}, \citenamefont {Kanzaki}, \citenamefont
  {Katori}, \citenamefont {Kawabata}, \citenamefont {Langanke}, \citenamefont
  {Mart\'{\i}nez-Pinedo}, \citenamefont {Noro}, \citenamefont {Roberts},
  \citenamefont {Sakaguchi}, \citenamefont {Shimbara}, \citenamefont {Shinada},
  \citenamefont {Stephenson}, \citenamefont {Ueno}, \citenamefont {Yamanaka},
  \citenamefont {Yoshifuku},\ and\ \citenamefont {Yosoi}}]{fuj07}%
  \BibitemOpen
  \bibfield  {author} {\bibinfo {author} {\bibfnamefont {H.}~\bibnamefont
  {Fujita}}, \bibinfo {author} {\bibfnamefont {Y.}~\bibnamefont {Fujita}},
  \bibinfo {author} {\bibfnamefont {T.}~\bibnamefont {Adachi}}, \bibinfo
  {author} {\bibfnamefont {A.}~\bibnamefont {Bacher}}, \bibinfo {author}
  {\bibfnamefont {G.}~\bibnamefont {Berg}}, \bibinfo {author} {\bibfnamefont
  {T.}~\bibnamefont {Black}}, \bibinfo {author} {\bibfnamefont
  {E.}~\bibnamefont {Caurier}}, \bibinfo {author} {\bibfnamefont
  {C.}~\bibnamefont {Foster}}, \bibinfo {author} {\bibfnamefont
  {H.}~\bibnamefont {Fujimura}}, \bibinfo {author} {\bibfnamefont
  {K.}~\bibnamefont {Hara}}, \bibinfo {author} {\bibfnamefont {K.}~\bibnamefont
  {Harada}}, \bibinfo {author} {\bibfnamefont {K.}~\bibnamefont {Hatanaka}},
  \bibinfo {author} {\bibfnamefont {J.}~\bibnamefont {J\"{a}necke}}, \bibinfo
  {author} {\bibfnamefont {J.}~\bibnamefont {Kamiya}}, \bibinfo {author}
  {\bibfnamefont {Y.}~\bibnamefont {Kanzaki}}, \bibinfo {author} {\bibfnamefont
  {K.}~\bibnamefont {Katori}}, \bibinfo {author} {\bibfnamefont
  {T.}~\bibnamefont {Kawabata}}, \bibinfo {author} {\bibfnamefont
  {K.}~\bibnamefont {Langanke}}, \bibinfo {author} {\bibfnamefont
  {G.}~\bibnamefont {Mart\'{\i}nez-Pinedo}}, \bibinfo {author} {\bibfnamefont
  {T.}~\bibnamefont {Noro}}, \bibinfo {author} {\bibfnamefont {D.}~\bibnamefont
  {Roberts}}, \bibinfo {author} {\bibfnamefont {H.}~\bibnamefont {Sakaguchi}},
  \bibinfo {author} {\bibfnamefont {Y.}~\bibnamefont {Shimbara}}, \bibinfo
  {author} {\bibfnamefont {T.}~\bibnamefont {Shinada}}, \bibinfo {author}
  {\bibfnamefont {E.}~\bibnamefont {Stephenson}}, \bibinfo {author}
  {\bibfnamefont {H.}~\bibnamefont {Ueno}}, \bibinfo {author} {\bibfnamefont
  {T.}~\bibnamefont {Yamanaka}}, \bibinfo {author} {\bibfnamefont
  {M.}~\bibnamefont {Yoshifuku}},\ and\ \bibinfo {author} {\bibfnamefont
  {M.}~\bibnamefont {Yosoi}},\ }\bibfield  {title} {\bibinfo {title} {Isospin
  structure of ${J}^\pi = 1^+$ states in $^{58}\mathrm{Ni}$ and
  $^{58}\mathrm{Cu}$ studied by $^{58}\mathrm{Ni}(p,p^\prime)$ and
  $^{58}\mathrm{Ni}(^3\mathrm{He},t)^{58}\mathrm{Cu}$ measurements},\
  }\href@noop {} {\bibfield  {journal} {\bibinfo  {journal} {Phys. Rev. C}\
  }\textbf {\bibinfo {volume} {75}},\ \bibinfo {pages} {034310} (\bibinfo
  {year} {2007})}\BibitemShut {NoStop}%
\bibitem [{\citenamefont {Hagemann}\ \emph {et~al.}(2005)\citenamefont
  {Hagemann}, \citenamefont {B\"aumer}, \citenamefont {van~den Berg},
  \citenamefont {De~Frenne}, \citenamefont {Frekers}, \citenamefont {Hannen},
  \citenamefont {Harakeh}, \citenamefont {Heyse}, \citenamefont {de~Huu},
  \citenamefont {Jacobs}, \citenamefont {Langanke}, \citenamefont
  {Mart\'{\i}nez-Pinedo}, \citenamefont {Negret}, \citenamefont {Popescu},
  \citenamefont {Rakers}, \citenamefont {Schmidt},\ and\ \citenamefont
  {W\"ortche}}]{hag05}%
  \BibitemOpen
  \bibfield  {author} {\bibinfo {author} {\bibfnamefont {M.}~\bibnamefont
  {Hagemann}}, \bibinfo {author} {\bibfnamefont {C.}~\bibnamefont {B\"aumer}},
  \bibinfo {author} {\bibfnamefont {A.~M.}\ \bibnamefont {van~den Berg}},
  \bibinfo {author} {\bibfnamefont {D.}~\bibnamefont {De~Frenne}}, \bibinfo
  {author} {\bibfnamefont {D.}~\bibnamefont {Frekers}}, \bibinfo {author}
  {\bibfnamefont {V.~M.}\ \bibnamefont {Hannen}}, \bibinfo {author}
  {\bibfnamefont {M.~N.}\ \bibnamefont {Harakeh}}, \bibinfo {author}
  {\bibfnamefont {J.}~\bibnamefont {Heyse}}, \bibinfo {author} {\bibfnamefont
  {M.~A.}\ \bibnamefont {de~Huu}}, \bibinfo {author} {\bibfnamefont
  {E.}~\bibnamefont {Jacobs}}, \bibinfo {author} {\bibfnamefont
  {K.}~\bibnamefont {Langanke}}, \bibinfo {author} {\bibfnamefont
  {G.}~\bibnamefont {Mart\'{\i}nez-Pinedo}}, \bibinfo {author} {\bibfnamefont
  {A.}~\bibnamefont {Negret}}, \bibinfo {author} {\bibfnamefont
  {L.}~\bibnamefont {Popescu}}, \bibinfo {author} {\bibfnamefont
  {S.}~\bibnamefont {Rakers}}, \bibinfo {author} {\bibfnamefont
  {R.}~\bibnamefont {Schmidt}},\ and\ \bibinfo {author} {\bibfnamefont {H.~J.}\
  \bibnamefont {W\"ortche}},\ }\bibfield  {title} {\bibinfo {title}
  {Spin-isospin excitations in the medium-mass nucleus $^{58}\mathrm{Co}$
  investigated with the ($d,{}^2\mathrm{He})$ reaction},\ }\href
  {https://doi.org/10.1103/PhysRevC.71.014606} {\bibfield  {journal} {\bibinfo
  {journal} {Phys. Rev. C}\ }\textbf {\bibinfo {volume} {71}},\ \bibinfo
  {pages} {014606} (\bibinfo {year} {2005})}\BibitemShut {NoStop}%
\bibitem [{\citenamefont {von Neumann-Cosel}\ \emph {et~al.}()\citenamefont
  {von Neumann-Cosel}, \citenamefont {Nesterenko}, \citenamefont {Brandherm},
  \citenamefont {Vishnevskiy}, \citenamefont {Reinhard}, \citenamefont
  {Kvasil}, \citenamefont {Matsubara}, \citenamefont {Repko}, \citenamefont
  {Richter}, \citenamefont {Scheck},\ and\ \citenamefont {Tamii}}]{vnc23}%
  \BibitemOpen
  \bibfield  {author} {\bibinfo {author} {\bibfnamefont {P.}~\bibnamefont {von
  Neumann-Cosel}}, \bibinfo {author} {\bibfnamefont {V.~O.}\ \bibnamefont
  {Nesterenko}}, \bibinfo {author} {\bibfnamefont {I.}~\bibnamefont
  {Brandherm}}, \bibinfo {author} {\bibfnamefont {P.~I.}\ \bibnamefont
  {Vishnevskiy}}, \bibinfo {author} {\bibfnamefont {P.~G.}\ \bibnamefont
  {Reinhard}}, \bibinfo {author} {\bibfnamefont {J.}~\bibnamefont {Kvasil}},
  \bibinfo {author} {\bibfnamefont {H.}~\bibnamefont {Matsubara}}, \bibinfo
  {author} {\bibfnamefont {A.}~\bibnamefont {Repko}}, \bibinfo {author}
  {\bibfnamefont {A.}~\bibnamefont {Richter}}, \bibinfo {author} {\bibfnamefont
  {M.}~\bibnamefont {Scheck}},\ and\ \bibinfo {author} {\bibfnamefont
  {A.}~\bibnamefont {Tamii}},\ }\href@noop {} {\bibinfo {title} {Evidence for a
  toroidal electric dipole mode in nuclei}},\ \Eprint
  {https://arxiv.org/abs/2310.04736} {arXiv:2310.04736 [nucl-ex]} \BibitemShut
  {NoStop}%
\bibitem [{\citenamefont {Repko}\ \emph {et~al.}(2019)\citenamefont {Repko},
  \citenamefont {Nesterenko}, \citenamefont {Kvasil},\ and\ \citenamefont
  {Reinhard}}]{rep19}%
  \BibitemOpen
  \bibfield  {author} {\bibinfo {author} {\bibfnamefont {A.}~\bibnamefont
  {Repko}}, \bibinfo {author} {\bibfnamefont {V.~O.}\ \bibnamefont
  {Nesterenko}}, \bibinfo {author} {\bibfnamefont {J.}~\bibnamefont {Kvasil}},\
  and\ \bibinfo {author} {\bibfnamefont {P.-G.}\ \bibnamefont {Reinhard}},\
  }\bibfield  {title} {\bibinfo {title} {Systematics of toroidal dipole modes
  in {C}a, {N}i, {Z}r, and {S}n isotopes},\ }\href
  {https://doi.org/10.1140/epja/i2019-12770-x} {\bibfield  {journal} {\bibinfo
  {journal} {Eur. Phys. J. A}\ }\textbf {\bibinfo {volume} {55}},\ \bibinfo
  {pages} {242} (\bibinfo {year} {2019})}\BibitemShut {NoStop}%
\bibitem [{\citenamefont {{Y. Fujita}}\ \emph {et~al.}(2002)\citenamefont {{Y.
  Fujita}}, \citenamefont {{H. Fujita}}, \citenamefont {{T. Adachi}},
  \citenamefont {{G. P. A. Berg}}, \citenamefont {{E. Caurier}}, \citenamefont
  {{H. Fujimura}}, \citenamefont {{K. Hara}}, \citenamefont {{K. Hatanaka}},
  \citenamefont {{Z. Janas}}, \citenamefont {{J. Kamiya}}, \citenamefont {{T.
  Kawabata}}, \citenamefont {{K. Langanke}}, \citenamefont {{G.
  Martínez-Pinedo}}, \citenamefont {{T. Noro}}, \citenamefont {{E. Roeckl}},
  \citenamefont {{Y. Shimbara}}, \citenamefont {{T. Shinada}}, \citenamefont
  {{S. Y. van der Werf}}, \citenamefont {{M. Yoshifuku}}, \citenamefont {{M.
  Yosoi}},\ and\ \citenamefont {{R. G. T. Zegers}}}]{fuj02}%
  \BibitemOpen
  \bibfield  {author} {\bibinfo {author} {\bibnamefont {{Y. Fujita}}}, \bibinfo
  {author} {\bibnamefont {{H. Fujita}}}, \bibinfo {author} {\bibnamefont {{T.
  Adachi}}}, \bibinfo {author} {\bibnamefont {{G. P. A. Berg}}}, \bibinfo
  {author} {\bibnamefont {{E. Caurier}}}, \bibinfo {author} {\bibnamefont {{H.
  Fujimura}}}, \bibinfo {author} {\bibnamefont {{K. Hara}}}, \bibinfo {author}
  {\bibnamefont {{K. Hatanaka}}}, \bibinfo {author} {\bibnamefont {{Z.
  Janas}}}, \bibinfo {author} {\bibnamefont {{J. Kamiya}}}, \bibinfo {author}
  {\bibnamefont {{T. Kawabata}}}, \bibinfo {author} {\bibnamefont {{K.
  Langanke}}}, \bibinfo {author} {\bibnamefont {{G. Martínez-Pinedo}}},
  \bibinfo {author} {\bibnamefont {{T. Noro}}}, \bibinfo {author} {\bibnamefont
  {{E. Roeckl}}}, \bibinfo {author} {\bibnamefont {{Y. Shimbara}}}, \bibinfo
  {author} {\bibnamefont {{T. Shinada}}}, \bibinfo {author} {\bibnamefont {{S.
  Y. van der Werf}}}, \bibinfo {author} {\bibnamefont {{M. Yoshifuku}}},
  \bibinfo {author} {\bibnamefont {{M. Yosoi}}},\ and\ \bibinfo {author}
  {\bibnamefont {{R. G. T. Zegers}}},\ }\bibfield  {title} {\bibinfo {title}
  {{G}amow-{T}eller transitions from $^{58}\mathrm{Ni}$ to discrete states of
  $^{58}\mathrm{Cu}$ - {T}he study of isospin symmetry in atomic nuclei},\
  }\href {https://doi.org/10.1140/epja/iepja1344} {\bibfield  {journal}
  {\bibinfo  {journal} {Eur. Phys. J. A}\ }\textbf {\bibinfo {volume} {13}},\
  \bibinfo {pages} {411} (\bibinfo {year} {2002})}\BibitemShut {NoStop}%
\bibitem [{\citenamefont {von Neumann-Cosel}\ and\ \citenamefont
  {Tamii}(2019)}]{vnc19}%
  \BibitemOpen
  \bibfield  {author} {\bibinfo {author} {\bibfnamefont {P.}~\bibnamefont {von
  Neumann-Cosel}}\ and\ \bibinfo {author} {\bibfnamefont {A.}~\bibnamefont
  {Tamii}},\ }\bibfield  {title} {\bibinfo {title} {Electric and magnetic
  dipole modes in high-resolution inelastic proton scattering at
  0{\textdegree}},\ }\href@noop {} {\bibfield  {journal} {\bibinfo  {journal}
  {Eur. Phys. J. A}\ }\textbf {\bibinfo {volume} {55}},\ \bibinfo {pages} {110}
  (\bibinfo {year} {2019})}\BibitemShut {NoStop}%
\bibitem [{\citenamefont {Tamii}\ \emph {et~al.}(2009)\citenamefont {Tamii},
  \citenamefont {Fujita}, \citenamefont {Matsubara}, \citenamefont {Adachi},
  \citenamefont {Carter}, \citenamefont {Dozono}, \citenamefont {Fujita},
  \citenamefont {Fujita}, \citenamefont {Hashimoto}, \citenamefont {Hatanaka},
  \citenamefont {Itahashi}, \citenamefont {Itoh}, \citenamefont {Kawabata},
  \citenamefont {Nakanishi}, \citenamefont {Ninomiya}, \citenamefont
  {Perez-Cerdan}, \citenamefont {Popescu}, \citenamefont {Rubio}, \citenamefont
  {Saito}, \citenamefont {Sakaguchi}, \citenamefont {Sakemi}, \citenamefont
  {Sasamoto}, \citenamefont {Shimbara}, \citenamefont {Shimizu}, \citenamefont
  {Smit}, \citenamefont {Tameshige}, \citenamefont {Yosoi},\ and\ \citenamefont
  {Zenhiro}}]{tam09}%
  \BibitemOpen
  \bibfield  {author} {\bibinfo {author} {\bibfnamefont {A.}~\bibnamefont
  {Tamii}}, \bibinfo {author} {\bibfnamefont {Y.}~\bibnamefont {Fujita}},
  \bibinfo {author} {\bibfnamefont {H.}~\bibnamefont {Matsubara}}, \bibinfo
  {author} {\bibfnamefont {T.}~\bibnamefont {Adachi}}, \bibinfo {author}
  {\bibfnamefont {J.}~\bibnamefont {Carter}}, \bibinfo {author} {\bibfnamefont
  {M.}~\bibnamefont {Dozono}}, \bibinfo {author} {\bibfnamefont
  {H.}~\bibnamefont {Fujita}}, \bibinfo {author} {\bibfnamefont
  {K.}~\bibnamefont {Fujita}}, \bibinfo {author} {\bibfnamefont
  {H.}~\bibnamefont {Hashimoto}}, \bibinfo {author} {\bibfnamefont
  {K.}~\bibnamefont {Hatanaka}}, \bibinfo {author} {\bibfnamefont
  {T.}~\bibnamefont {Itahashi}}, \bibinfo {author} {\bibfnamefont
  {M.}~\bibnamefont {Itoh}}, \bibinfo {author} {\bibfnamefont {T.}~\bibnamefont
  {Kawabata}}, \bibinfo {author} {\bibfnamefont {K.}~\bibnamefont {Nakanishi}},
  \bibinfo {author} {\bibfnamefont {S.}~\bibnamefont {Ninomiya}}, \bibinfo
  {author} {\bibfnamefont {A.}~\bibnamefont {Perez-Cerdan}}, \bibinfo {author}
  {\bibfnamefont {L.}~\bibnamefont {Popescu}}, \bibinfo {author} {\bibfnamefont
  {B.}~\bibnamefont {Rubio}}, \bibinfo {author} {\bibfnamefont
  {T.}~\bibnamefont {Saito}}, \bibinfo {author} {\bibfnamefont
  {H.}~\bibnamefont {Sakaguchi}}, \bibinfo {author} {\bibfnamefont
  {Y.}~\bibnamefont {Sakemi}}, \bibinfo {author} {\bibfnamefont
  {Y.}~\bibnamefont {Sasamoto}}, \bibinfo {author} {\bibfnamefont
  {Y.}~\bibnamefont {Shimbara}}, \bibinfo {author} {\bibfnamefont
  {Y.}~\bibnamefont {Shimizu}}, \bibinfo {author} {\bibfnamefont
  {F.}~\bibnamefont {Smit}}, \bibinfo {author} {\bibfnamefont {Y.}~\bibnamefont
  {Tameshige}}, \bibinfo {author} {\bibfnamefont {M.}~\bibnamefont {Yosoi}},\
  and\ \bibinfo {author} {\bibfnamefont {J.}~\bibnamefont {Zenhiro}},\
  }\bibfield  {title} {\bibinfo {title} {Measurement of high energy resolution
  inelastic proton scattering at and close to zero degrees},\ }\href
  {https://doi.org/https://doi.org/10.1016/j.nima.2009.03.248} {\bibfield
  {journal} {\bibinfo  {journal} {Nucl. Instrum. Methods Phys. Res., Sect. A}\
  }\textbf {\bibinfo {volume} {605}},\ \bibinfo {pages} {326} (\bibinfo {year}
  {2009})}\BibitemShut {NoStop}%
\bibitem [{\citenamefont {Mathy}\ \emph {et~al.}(2017)\citenamefont {Mathy},
  \citenamefont {Birkhan}, \citenamefont {Matsubara}, \citenamefont {von
  Neumann-Cosel}, \citenamefont {Pietralla}, \citenamefont {Ponomarev},
  \citenamefont {Richter},\ and\ \citenamefont {Tamii}}]{mat17}%
  \BibitemOpen
  \bibfield  {author} {\bibinfo {author} {\bibfnamefont {M.}~\bibnamefont
  {Mathy}}, \bibinfo {author} {\bibfnamefont {J.}~\bibnamefont {Birkhan}},
  \bibinfo {author} {\bibfnamefont {H.}~\bibnamefont {Matsubara}}, \bibinfo
  {author} {\bibfnamefont {P.}~\bibnamefont {von Neumann-Cosel}}, \bibinfo
  {author} {\bibfnamefont {N.}~\bibnamefont {Pietralla}}, \bibinfo {author}
  {\bibfnamefont {V.~Y.}\ \bibnamefont {Ponomarev}}, \bibinfo {author}
  {\bibfnamefont {A.}~\bibnamefont {Richter}},\ and\ \bibinfo {author}
  {\bibfnamefont {A.}~\bibnamefont {Tamii}},\ }\bibfield  {title} {\bibinfo
  {title} {Search for weak ${M}1$ transitions in $^{48}\mathrm{Ca}$ with
  inelastic proton scattering},\ }\href
  {https://doi.org/10.1103/PhysRevC.95.054316} {\bibfield  {journal} {\bibinfo
  {journal} {Phys. Rev. C}\ }\textbf {\bibinfo {volume} {95}},\ \bibinfo
  {pages} {054316} (\bibinfo {year} {2017})}\BibitemShut {NoStop}%
\bibitem [{\citenamefont {Fujiwara}\ \emph {et~al.}(1999)\citenamefont
  {Fujiwara}, \citenamefont {Akimune}, \citenamefont {Daito}, \citenamefont
  {Fujimura}, \citenamefont {Fujita}, \citenamefont {Hatanaka}, \citenamefont
  {Ikegami}, \citenamefont {Katayama}, \citenamefont {Nagayama}, \citenamefont
  {Matsuoka}, \citenamefont {Morinobu}, \citenamefont {Noro}, \citenamefont
  {Yoshimura}, \citenamefont {Sakaguchi}, \citenamefont {Sakemi}, \citenamefont
  {Tamii},\ and\ \citenamefont {Yosoi}}]{fuj99}%
  \BibitemOpen
  \bibfield  {author} {\bibinfo {author} {\bibfnamefont {M.}~\bibnamefont
  {Fujiwara}}, \bibinfo {author} {\bibfnamefont {H.}~\bibnamefont {Akimune}},
  \bibinfo {author} {\bibfnamefont {I.}~\bibnamefont {Daito}}, \bibinfo
  {author} {\bibfnamefont {H.}~\bibnamefont {Fujimura}}, \bibinfo {author}
  {\bibfnamefont {Y.}~\bibnamefont {Fujita}}, \bibinfo {author} {\bibfnamefont
  {K.}~\bibnamefont {Hatanaka}}, \bibinfo {author} {\bibfnamefont
  {H.}~\bibnamefont {Ikegami}}, \bibinfo {author} {\bibfnamefont
  {I.}~\bibnamefont {Katayama}}, \bibinfo {author} {\bibfnamefont
  {K.}~\bibnamefont {Nagayama}}, \bibinfo {author} {\bibfnamefont
  {N.}~\bibnamefont {Matsuoka}}, \bibinfo {author} {\bibfnamefont
  {S.}~\bibnamefont {Morinobu}}, \bibinfo {author} {\bibfnamefont
  {T.}~\bibnamefont {Noro}}, \bibinfo {author} {\bibfnamefont {M.}~\bibnamefont
  {Yoshimura}}, \bibinfo {author} {\bibfnamefont {H.}~\bibnamefont
  {Sakaguchi}}, \bibinfo {author} {\bibfnamefont {Y.}~\bibnamefont {Sakemi}},
  \bibinfo {author} {\bibfnamefont {A.}~\bibnamefont {Tamii}},\ and\ \bibinfo
  {author} {\bibfnamefont {M.}~\bibnamefont {Yosoi}},\ }\bibfield  {title}
  {\bibinfo {title} {Magnetic spectrometer {G}rand {R}aiden},\ }\href
  {https://doi.org/https://doi.org/10.1016/S0168-9002(98)01009-2} {\bibfield
  {journal} {\bibinfo  {journal} {Nucl. Instrum. Methods Phys. Res., Sect. A}\
  }\textbf {\bibinfo {volume} {422}},\ \bibinfo {pages} {484} (\bibinfo {year}
  {1999})}\BibitemShut {NoStop}%
\bibitem [{\citenamefont {Birkhan}\ \emph {et~al.}(2017)\citenamefont
  {Birkhan}, \citenamefont {Miorelli}, \citenamefont {Bacca}, \citenamefont
  {Bassauer}, \citenamefont {Bertulani}, \citenamefont {Hagen}, \citenamefont
  {Matsubara}, \citenamefont {von Neumann-Cosel}, \citenamefont {Papenbrock},
  \citenamefont {Pietralla}, \citenamefont {Ponomarev}, \citenamefont
  {Richter}, \citenamefont {Schwenk},\ and\ \citenamefont {Tamii}}]{bir17}%
  \BibitemOpen
  \bibfield  {author} {\bibinfo {author} {\bibfnamefont {J.}~\bibnamefont
  {Birkhan}}, \bibinfo {author} {\bibfnamefont {M.}~\bibnamefont {Miorelli}},
  \bibinfo {author} {\bibfnamefont {S.}~\bibnamefont {Bacca}}, \bibinfo
  {author} {\bibfnamefont {S.}~\bibnamefont {Bassauer}}, \bibinfo {author}
  {\bibfnamefont {C.~A.}\ \bibnamefont {Bertulani}}, \bibinfo {author}
  {\bibfnamefont {G.}~\bibnamefont {Hagen}}, \bibinfo {author} {\bibfnamefont
  {H.}~\bibnamefont {Matsubara}}, \bibinfo {author} {\bibfnamefont
  {P.}~\bibnamefont {von Neumann-Cosel}}, \bibinfo {author} {\bibfnamefont
  {T.}~\bibnamefont {Papenbrock}}, \bibinfo {author} {\bibfnamefont
  {N.}~\bibnamefont {Pietralla}}, \bibinfo {author} {\bibfnamefont {V.~Y.}\
  \bibnamefont {Ponomarev}}, \bibinfo {author} {\bibfnamefont {A.}~\bibnamefont
  {Richter}}, \bibinfo {author} {\bibfnamefont {A.}~\bibnamefont {Schwenk}},\
  and\ \bibinfo {author} {\bibfnamefont {A.}~\bibnamefont {Tamii}},\ }\bibfield
   {title} {\bibinfo {title} {Electric dipole polarizability of
  $^{48}\mathrm{Ca}$ and implications for the neutron skin},\ }\href
  {https://doi.org/10.1103/PhysRevLett.118.252501} {\bibfield  {journal}
  {\bibinfo  {journal} {Phys. Rev. Lett.}\ }\textbf {\bibinfo {volume} {118}},\
  \bibinfo {pages} {252501} (\bibinfo {year} {2017})}\BibitemShut {NoStop}%
\bibitem [{\citenamefont {Bassauer}\ \emph {et~al.}(2020)\citenamefont
  {Bassauer}, \citenamefont {von Neumann-Cosel}, \citenamefont {Reinhard},
  \citenamefont {Tamii}, \citenamefont {Adachi}, \citenamefont {Bertulani},
  \citenamefont {Chan}, \citenamefont {D'Alessio}, \citenamefont {Fujioka},
  \citenamefont {Fujita}, \citenamefont {Fujita}, \citenamefont {Gey},
  \citenamefont {Hilcker}, \citenamefont {Hoang}, \citenamefont {Inoue},
  \citenamefont {Isaak}, \citenamefont {Iwamoto}, \citenamefont {Klaus},
  \citenamefont {Kobayashi}, \citenamefont {Maeda}, \citenamefont {Matsuda},
  \citenamefont {Nakatsuka}, \citenamefont {Noji}, \citenamefont {Ong},
  \citenamefont {Ou}, \citenamefont {Pietralla}, \citenamefont {Ponomarev},
  \citenamefont {Reen}, \citenamefont {Richter}, \citenamefont {Singer},
  \citenamefont {Steinhilber}, \citenamefont {Sudo}, \citenamefont {Togano},
  \citenamefont {Tsumura}, \citenamefont {Watanabe},\ and\ \citenamefont
  {Werner}}]{bas20}%
  \BibitemOpen
  \bibfield  {author} {\bibinfo {author} {\bibfnamefont {S.}~\bibnamefont
  {Bassauer}}, \bibinfo {author} {\bibfnamefont {P.}~\bibnamefont {von
  Neumann-Cosel}}, \bibinfo {author} {\bibfnamefont {P.-G.}\ \bibnamefont
  {Reinhard}}, \bibinfo {author} {\bibfnamefont {A.}~\bibnamefont {Tamii}},
  \bibinfo {author} {\bibfnamefont {S.}~\bibnamefont {Adachi}}, \bibinfo
  {author} {\bibfnamefont {C.~A.}\ \bibnamefont {Bertulani}}, \bibinfo {author}
  {\bibfnamefont {P.~Y.}\ \bibnamefont {Chan}}, \bibinfo {author}
  {\bibfnamefont {A.}~\bibnamefont {D'Alessio}}, \bibinfo {author}
  {\bibfnamefont {H.}~\bibnamefont {Fujioka}}, \bibinfo {author} {\bibfnamefont
  {H.}~\bibnamefont {Fujita}}, \bibinfo {author} {\bibfnamefont
  {Y.}~\bibnamefont {Fujita}}, \bibinfo {author} {\bibfnamefont
  {G.}~\bibnamefont {Gey}}, \bibinfo {author} {\bibfnamefont {M.}~\bibnamefont
  {Hilcker}}, \bibinfo {author} {\bibfnamefont {T.~H.}\ \bibnamefont {Hoang}},
  \bibinfo {author} {\bibfnamefont {A.}~\bibnamefont {Inoue}}, \bibinfo
  {author} {\bibfnamefont {J.}~\bibnamefont {Isaak}}, \bibinfo {author}
  {\bibfnamefont {C.}~\bibnamefont {Iwamoto}}, \bibinfo {author} {\bibfnamefont
  {T.}~\bibnamefont {Klaus}}, \bibinfo {author} {\bibfnamefont
  {N.}~\bibnamefont {Kobayashi}}, \bibinfo {author} {\bibfnamefont
  {Y.}~\bibnamefont {Maeda}}, \bibinfo {author} {\bibfnamefont
  {M.}~\bibnamefont {Matsuda}}, \bibinfo {author} {\bibfnamefont
  {N.}~\bibnamefont {Nakatsuka}}, \bibinfo {author} {\bibfnamefont
  {S.}~\bibnamefont {Noji}}, \bibinfo {author} {\bibfnamefont {H.~J.}\
  \bibnamefont {Ong}}, \bibinfo {author} {\bibfnamefont {I.}~\bibnamefont
  {Ou}}, \bibinfo {author} {\bibfnamefont {N.}~\bibnamefont {Pietralla}},
  \bibinfo {author} {\bibfnamefont {V.~Y.}\ \bibnamefont {Ponomarev}}, \bibinfo
  {author} {\bibfnamefont {M.~S.}\ \bibnamefont {Reen}}, \bibinfo {author}
  {\bibfnamefont {A.}~\bibnamefont {Richter}}, \bibinfo {author} {\bibfnamefont
  {M.}~\bibnamefont {Singer}}, \bibinfo {author} {\bibfnamefont
  {G.}~\bibnamefont {Steinhilber}}, \bibinfo {author} {\bibfnamefont
  {T.}~\bibnamefont {Sudo}}, \bibinfo {author} {\bibfnamefont {Y.}~\bibnamefont
  {Togano}}, \bibinfo {author} {\bibfnamefont {M.}~\bibnamefont {Tsumura}},
  \bibinfo {author} {\bibfnamefont {Y.}~\bibnamefont {Watanabe}},\ and\
  \bibinfo {author} {\bibfnamefont {V.}~\bibnamefont {Werner}},\ }\bibfield
  {title} {\bibinfo {title} {Electric and magnetic dipole strength in
  $^{112,114,116,118,120,124}\mathrm{Sn}$},\ }\href
  {https://doi.org/10.1103/PhysRevC.102.034327} {\bibfield  {journal} {\bibinfo
   {journal} {Phys. Rev. C}\ }\textbf {\bibinfo {volume} {102}},\ \bibinfo
  {pages} {034327} (\bibinfo {year} {2020})}\BibitemShut {NoStop}%
\bibitem [{\citenamefont {Fearick}\ \emph {et~al.}(2023)\citenamefont
  {Fearick}, \citenamefont {von Neumann-Cosel}, \citenamefont {Bacca},
  \citenamefont {Birkhan}, \citenamefont {Bonaiti}, \citenamefont {Brandherm},
  \citenamefont {Hagen}, \citenamefont {Matsubara}, \citenamefont {Nazarewicz},
  \citenamefont {Pietralla}, \citenamefont {Ponomarev}, \citenamefont
  {Reinhard}, \citenamefont {Roca-Maza}, \citenamefont {Richter}, \citenamefont
  {Schwenk}, \citenamefont {Simonis},\ and\ \citenamefont {Tamii}}]{fea23}%
  \BibitemOpen
  \bibfield  {author} {\bibinfo {author} {\bibfnamefont {R.~W.}\ \bibnamefont
  {Fearick}}, \bibinfo {author} {\bibfnamefont {P.}~\bibnamefont {von
  Neumann-Cosel}}, \bibinfo {author} {\bibfnamefont {S.}~\bibnamefont {Bacca}},
  \bibinfo {author} {\bibfnamefont {J.}~\bibnamefont {Birkhan}}, \bibinfo
  {author} {\bibfnamefont {F.}~\bibnamefont {Bonaiti}}, \bibinfo {author}
  {\bibfnamefont {I.}~\bibnamefont {Brandherm}}, \bibinfo {author}
  {\bibfnamefont {G.}~\bibnamefont {Hagen}}, \bibinfo {author} {\bibfnamefont
  {H.}~\bibnamefont {Matsubara}}, \bibinfo {author} {\bibfnamefont
  {W.}~\bibnamefont {Nazarewicz}}, \bibinfo {author} {\bibfnamefont
  {N.}~\bibnamefont {Pietralla}}, \bibinfo {author} {\bibfnamefont {V.~Y.}\
  \bibnamefont {Ponomarev}}, \bibinfo {author} {\bibfnamefont {P.-G.}\
  \bibnamefont {Reinhard}}, \bibinfo {author} {\bibfnamefont {X.}~\bibnamefont
  {Roca-Maza}}, \bibinfo {author} {\bibfnamefont {A.}~\bibnamefont {Richter}},
  \bibinfo {author} {\bibfnamefont {A.}~\bibnamefont {Schwenk}}, \bibinfo
  {author} {\bibfnamefont {J.}~\bibnamefont {Simonis}},\ and\ \bibinfo {author}
  {\bibfnamefont {A.}~\bibnamefont {Tamii}},\ }\bibfield  {title} {\bibinfo
  {title} {Electric dipole polarizability of $^{40}\mathrm{Ca}$},\ }\href
  {https://doi.org/10.1103/PhysRevResearch.5.L022044} {\bibfield  {journal}
  {\bibinfo  {journal} {Phys. Rev. Res.}\ }\textbf {\bibinfo {volume} {5}},\
  \bibinfo {pages} {L022044} (\bibinfo {year} {2023})}\BibitemShut {NoStop}%
\bibitem [{HDT()}]{HDTV}%
  \BibitemOpen
  \href@noop {} {}\bibinfo {howpublished} {Computer Code HDTV,
  https://gitlab.ikp.uni-koeln.de/staging/hdtv}\BibitemShut {NoStop}%
\bibitem [{\citenamefont {Raynal}(2007)}]{ray07}%
  \BibitemOpen
  \bibfield  {author} {\bibinfo {author} {\bibfnamefont {J.}~\bibnamefont
  {Raynal}},\ }\href@noop {} {}\bibinfo {howpublished} {{Computer Code: DWBA07,
  NEA Data Service NEA1209/008}} (\bibinfo {year} {2007})\BibitemShut {NoStop}%
\bibitem [{\citenamefont {Love}\ and\ \citenamefont {Franey}(1981)}]{lov81}%
  \BibitemOpen
  \bibfield  {author} {\bibinfo {author} {\bibfnamefont {W.~G.}\ \bibnamefont
  {Love}}\ and\ \bibinfo {author} {\bibfnamefont {M.~A.}\ \bibnamefont
  {Franey}},\ }\bibfield  {title} {\bibinfo {title} {Effective nucleon-nucleon
  interaction for scattering at intermediate energies},\ }\href
  {https://doi.org/10.1103/PhysRevC.24.1073} {\bibfield  {journal} {\bibinfo
  {journal} {Phys. Rev. C}\ }\textbf {\bibinfo {volume} {24}},\ \bibinfo
  {pages} {1073} (\bibinfo {year} {1981})}\BibitemShut {NoStop}%
\bibitem [{\citenamefont {Lui}\ \emph {et~al.}(2006)\citenamefont {Lui},
  \citenamefont {Youngblood}, \citenamefont {Clark}, \citenamefont {Tokimoto},\
  and\ \citenamefont {John}}]{lui06}%
  \BibitemOpen
  \bibfield  {author} {\bibinfo {author} {\bibfnamefont {Y.-W.}\ \bibnamefont
  {Lui}}, \bibinfo {author} {\bibfnamefont {D.~H.}\ \bibnamefont {Youngblood}},
  \bibinfo {author} {\bibfnamefont {H.~L.}\ \bibnamefont {Clark}}, \bibinfo
  {author} {\bibfnamefont {Y.}~\bibnamefont {Tokimoto}},\ and\ \bibinfo
  {author} {\bibfnamefont {B.}~\bibnamefont {John}},\ }\bibfield  {title}
  {\bibinfo {title} {Isoscalar giant resonances for nuclei with mass between 56
  and 60},\ }\href {https://doi.org/10.1103/PhysRevC.73.014314} {\bibfield
  {journal} {\bibinfo  {journal} {Phys. Rev. C}\ }\textbf {\bibinfo {volume}
  {73}},\ \bibinfo {pages} {014314} (\bibinfo {year} {2006})}\BibitemShut
  {NoStop}%
\bibitem [{\citenamefont {Taddeucci}\ \emph {et~al.}(1987)\citenamefont
  {Taddeucci}, \citenamefont {Goulding}, \citenamefont {Carey}, \citenamefont
  {Byrd}, \citenamefont {Goodman}, \citenamefont {Gaarde}, \citenamefont
  {Larsen}, \citenamefont {Horen}, \citenamefont {Rapaport},\ and\
  \citenamefont {Sugarbaker}}]{tad87}%
  \BibitemOpen
  \bibfield  {author} {\bibinfo {author} {\bibfnamefont {T.}~\bibnamefont
  {Taddeucci}}, \bibinfo {author} {\bibfnamefont {C.}~\bibnamefont {Goulding}},
  \bibinfo {author} {\bibfnamefont {T.}~\bibnamefont {Carey}}, \bibinfo
  {author} {\bibfnamefont {R.}~\bibnamefont {Byrd}}, \bibinfo {author}
  {\bibfnamefont {C.}~\bibnamefont {Goodman}}, \bibinfo {author} {\bibfnamefont
  {C.}~\bibnamefont {Gaarde}}, \bibinfo {author} {\bibfnamefont
  {J.}~\bibnamefont {Larsen}}, \bibinfo {author} {\bibfnamefont
  {D.}~\bibnamefont {Horen}}, \bibinfo {author} {\bibfnamefont
  {J.}~\bibnamefont {Rapaport}},\ and\ \bibinfo {author} {\bibfnamefont
  {E.}~\bibnamefont {Sugarbaker}},\ }\bibfield  {title} {\bibinfo {title} {The
  $(p,n)$ reaction as a probe of beta decay strength},\ }\href
  {https://doi.org/https://doi.org/10.1016/0375-9474(87)90089-3} {\bibfield
  {journal} {\bibinfo  {journal} {Nucl. Phys. A}\ }\textbf {\bibinfo {volume}
  {469}},\ \bibinfo {pages} {125} (\bibinfo {year} {1987})}\BibitemShut
  {NoStop}%
\bibitem [{\citenamefont {Sasano}\ \emph {et~al.}(2009)\citenamefont {Sasano},
  \citenamefont {Sakai}, \citenamefont {Yako}, \citenamefont {Wakasa},
  \citenamefont {Asaji}, \citenamefont {Fujita}, \citenamefont {Fujita},
  \citenamefont {Greenfield}, \citenamefont {Hagihara}, \citenamefont
  {Hatanaka}, \citenamefont {Kawabata}, \citenamefont {Kuboki}, \citenamefont
  {Maeda}, \citenamefont {Okamura}, \citenamefont {Saito}, \citenamefont
  {Sakemi}, \citenamefont {Sekiguchi}, \citenamefont {Shimizu}, \citenamefont
  {Takahashi}, \citenamefont {Tameshige},\ and\ \citenamefont {Tamii}}]{sas09}%
  \BibitemOpen
  \bibfield  {author} {\bibinfo {author} {\bibfnamefont {M.}~\bibnamefont
  {Sasano}}, \bibinfo {author} {\bibfnamefont {H.}~\bibnamefont {Sakai}},
  \bibinfo {author} {\bibfnamefont {K.}~\bibnamefont {Yako}}, \bibinfo {author}
  {\bibfnamefont {T.}~\bibnamefont {Wakasa}}, \bibinfo {author} {\bibfnamefont
  {S.}~\bibnamefont {Asaji}}, \bibinfo {author} {\bibfnamefont
  {K.}~\bibnamefont {Fujita}}, \bibinfo {author} {\bibfnamefont
  {Y.}~\bibnamefont {Fujita}}, \bibinfo {author} {\bibfnamefont {M.~B.}\
  \bibnamefont {Greenfield}}, \bibinfo {author} {\bibfnamefont
  {Y.}~\bibnamefont {Hagihara}}, \bibinfo {author} {\bibfnamefont
  {K.}~\bibnamefont {Hatanaka}}, \bibinfo {author} {\bibfnamefont
  {T.}~\bibnamefont {Kawabata}}, \bibinfo {author} {\bibfnamefont
  {H.}~\bibnamefont {Kuboki}}, \bibinfo {author} {\bibfnamefont
  {Y.}~\bibnamefont {Maeda}}, \bibinfo {author} {\bibfnamefont
  {H.}~\bibnamefont {Okamura}}, \bibinfo {author} {\bibfnamefont
  {T.}~\bibnamefont {Saito}}, \bibinfo {author} {\bibfnamefont
  {Y.}~\bibnamefont {Sakemi}}, \bibinfo {author} {\bibfnamefont
  {K.}~\bibnamefont {Sekiguchi}}, \bibinfo {author} {\bibfnamefont
  {Y.}~\bibnamefont {Shimizu}}, \bibinfo {author} {\bibfnamefont
  {Y.}~\bibnamefont {Takahashi}}, \bibinfo {author} {\bibfnamefont
  {Y.}~\bibnamefont {Tameshige}},\ and\ \bibinfo {author} {\bibfnamefont
  {A.}~\bibnamefont {Tamii}},\ }\bibfield  {title} {\bibinfo {title}
  {Gamow-{T}eller unit cross sections of the $(p,n)$ reaction at 198 and 297
  {M}ev on medium-heavy nuclei},\ }\href
  {https://doi.org/10.1103/PhysRevC.79.024602} {\bibfield  {journal} {\bibinfo
  {journal} {Phys. Rev. C}\ }\textbf {\bibinfo {volume} {79}},\ \bibinfo
  {pages} {024602} (\bibinfo {year} {2009})}\BibitemShut {NoStop}%
\bibitem [{\citenamefont {Birkhan}\ \emph {et~al.}(2016)\citenamefont
  {Birkhan}, \citenamefont {Matsubara}, \citenamefont {von Neumann-Cosel},
  \citenamefont {Pietralla}, \citenamefont {Ponomarev}, \citenamefont
  {Richter}, \citenamefont {Tamii},\ and\ \citenamefont {Wambach}}]{bir16}%
  \BibitemOpen
  \bibfield  {author} {\bibinfo {author} {\bibfnamefont {J.}~\bibnamefont
  {Birkhan}}, \bibinfo {author} {\bibfnamefont {H.}~\bibnamefont {Matsubara}},
  \bibinfo {author} {\bibfnamefont {P.}~\bibnamefont {von Neumann-Cosel}},
  \bibinfo {author} {\bibfnamefont {N.}~\bibnamefont {Pietralla}}, \bibinfo
  {author} {\bibfnamefont {V.~Y.}\ \bibnamefont {Ponomarev}}, \bibinfo {author}
  {\bibfnamefont {A.}~\bibnamefont {Richter}}, \bibinfo {author} {\bibfnamefont
  {A.}~\bibnamefont {Tamii}},\ and\ \bibinfo {author} {\bibfnamefont
  {J.}~\bibnamefont {Wambach}},\ }\bibfield  {title} {\bibinfo {title}
  {Electromagnetic ${M}1$ transition strengths from inelastic proton
  scattering: The cases of $^{48}\mathrm{Ca}$ and $^{208}\mathrm{Pb}$},\ }\href
  {https://doi.org/10.1103/PhysRevC.93.041302} {\bibfield  {journal} {\bibinfo
  {journal} {Phys. Rev. C}\ }\textbf {\bibinfo {volume} {93}},\ \bibinfo
  {pages} {041302(R)} (\bibinfo {year} {2016})}\BibitemShut {NoStop}%
\bibitem [{\citenamefont {Bertulani}\ and\ \citenamefont {Baur}(1988)}]{ber88}%
  \BibitemOpen
  \bibfield  {author} {\bibinfo {author} {\bibfnamefont {C.~A.}\ \bibnamefont
  {Bertulani}}\ and\ \bibinfo {author} {\bibfnamefont {G.}~\bibnamefont
  {Baur}},\ }\bibfield  {title} {\bibinfo {title} {Electromagnetic processes in
  relativistic heavy ion collisions},\ }\href
  {https://doi.org/https://doi.org/10.1016/0370-1573(88)90142-1} {\bibfield
  {journal} {\bibinfo  {journal} {Phys. Rep.}\ }\textbf {\bibinfo {volume}
  {163}},\ \bibinfo {pages} {299} (\bibinfo {year} {1988})}\BibitemShut
  {NoStop}%
\bibitem [{\citenamefont {Bertulani}\ and\ \citenamefont
  {Nathan}(1993)}]{ber93}%
  \BibitemOpen
  \bibfield  {author} {\bibinfo {author} {\bibfnamefont {C.}~\bibnamefont
  {Bertulani}}\ and\ \bibinfo {author} {\bibfnamefont {A.}~\bibnamefont
  {Nathan}},\ }\bibfield  {title} {\bibinfo {title} {Excitation and photon
  decay of giant resonances from high-energy collisions of heavy ions},\ }\href
  {https://doi.org/https://doi.org/10.1016/0375-9474(93)90363-3} {\bibfield
  {journal} {\bibinfo  {journal} {Nucl. Phys. A}\ }\textbf {\bibinfo {volume}
  {554}},\ \bibinfo {pages} {158} (\bibinfo {year} {1993})}\BibitemShut
  {NoStop}%
\bibitem [{\citenamefont {Poltoratska}\ \emph {et~al.}(2012)\citenamefont
  {Poltoratska}, \citenamefont {von Neumann-Cosel}, \citenamefont {Tamii},
  \citenamefont {Adachi}, \citenamefont {Bertulani}, \citenamefont {Carter},
  \citenamefont {Dozono}, \citenamefont {Fujita}, \citenamefont {Fujita},
  \citenamefont {Fujita}, \citenamefont {Hatanaka}, \citenamefont {Itoh},
  \citenamefont {Kawabata}, \citenamefont {Kalmykov}, \citenamefont
  {Krumbholz}, \citenamefont {Litvinova}, \citenamefont {Matsubara},
  \citenamefont {Nakanishi}, \citenamefont {Neveling}, \citenamefont {Okamura},
  \citenamefont {Ong}, \citenamefont {\"Ozel-Tashenov}, \citenamefont
  {Ponomarev}, \citenamefont {Richter}, \citenamefont {Rubio}, \citenamefont
  {Sakaguchi}, \citenamefont {Sakemi}, \citenamefont {Sasamoto}, \citenamefont
  {Shimbara}, \citenamefont {Shimizu}, \citenamefont {Smit}, \citenamefont
  {Suzuki}, \citenamefont {Tameshige}, \citenamefont {Wambach}, \citenamefont
  {Yosoi},\ and\ \citenamefont {Zenihiro}}]{pol12}%
  \BibitemOpen
  \bibfield  {author} {\bibinfo {author} {\bibfnamefont {I.}~\bibnamefont
  {Poltoratska}}, \bibinfo {author} {\bibfnamefont {P.}~\bibnamefont {von
  Neumann-Cosel}}, \bibinfo {author} {\bibfnamefont {A.}~\bibnamefont {Tamii}},
  \bibinfo {author} {\bibfnamefont {T.}~\bibnamefont {Adachi}}, \bibinfo
  {author} {\bibfnamefont {C.~A.}\ \bibnamefont {Bertulani}}, \bibinfo {author}
  {\bibfnamefont {J.}~\bibnamefont {Carter}}, \bibinfo {author} {\bibfnamefont
  {M.}~\bibnamefont {Dozono}}, \bibinfo {author} {\bibfnamefont
  {H.}~\bibnamefont {Fujita}}, \bibinfo {author} {\bibfnamefont
  {K.}~\bibnamefont {Fujita}}, \bibinfo {author} {\bibfnamefont
  {Y.}~\bibnamefont {Fujita}}, \bibinfo {author} {\bibfnamefont
  {K.}~\bibnamefont {Hatanaka}}, \bibinfo {author} {\bibfnamefont
  {M.}~\bibnamefont {Itoh}}, \bibinfo {author} {\bibfnamefont {T.}~\bibnamefont
  {Kawabata}}, \bibinfo {author} {\bibfnamefont {Y.}~\bibnamefont {Kalmykov}},
  \bibinfo {author} {\bibfnamefont {A.~M.}\ \bibnamefont {Krumbholz}}, \bibinfo
  {author} {\bibfnamefont {E.}~\bibnamefont {Litvinova}}, \bibinfo {author}
  {\bibfnamefont {H.}~\bibnamefont {Matsubara}}, \bibinfo {author}
  {\bibfnamefont {K.}~\bibnamefont {Nakanishi}}, \bibinfo {author}
  {\bibfnamefont {R.}~\bibnamefont {Neveling}}, \bibinfo {author}
  {\bibfnamefont {H.}~\bibnamefont {Okamura}}, \bibinfo {author} {\bibfnamefont
  {H.~J.}\ \bibnamefont {Ong}}, \bibinfo {author} {\bibfnamefont
  {B.}~\bibnamefont {\"Ozel-Tashenov}}, \bibinfo {author} {\bibfnamefont
  {V.~Y.}\ \bibnamefont {Ponomarev}}, \bibinfo {author} {\bibfnamefont
  {A.}~\bibnamefont {Richter}}, \bibinfo {author} {\bibfnamefont
  {B.}~\bibnamefont {Rubio}}, \bibinfo {author} {\bibfnamefont
  {H.}~\bibnamefont {Sakaguchi}}, \bibinfo {author} {\bibfnamefont
  {Y.}~\bibnamefont {Sakemi}}, \bibinfo {author} {\bibfnamefont
  {Y.}~\bibnamefont {Sasamoto}}, \bibinfo {author} {\bibfnamefont
  {Y.}~\bibnamefont {Shimbara}}, \bibinfo {author} {\bibfnamefont
  {Y.}~\bibnamefont {Shimizu}}, \bibinfo {author} {\bibfnamefont {F.~D.}\
  \bibnamefont {Smit}}, \bibinfo {author} {\bibfnamefont {T.}~\bibnamefont
  {Suzuki}}, \bibinfo {author} {\bibfnamefont {Y.}~\bibnamefont {Tameshige}},
  \bibinfo {author} {\bibfnamefont {J.}~\bibnamefont {Wambach}}, \bibinfo
  {author} {\bibfnamefont {M.}~\bibnamefont {Yosoi}},\ and\ \bibinfo {author}
  {\bibfnamefont {J.}~\bibnamefont {Zenihiro}},\ }\bibfield  {title} {\bibinfo
  {title} {Pygmy dipole resonance in ${}^{208}${P}b},\ }\href
  {https://doi.org/10.1103/PhysRevC.85.041304} {\bibfield  {journal} {\bibinfo
  {journal} {Phys. Rev. C}\ }\textbf {\bibinfo {volume} {85}},\ \bibinfo
  {pages} {041304} (\bibinfo {year} {2012})}\BibitemShut {NoStop}%
\bibitem [{\citenamefont {Weise}\ and\ \citenamefont {Wogner}(1995)}]{wei94}%
  \BibitemOpen
  \bibfield  {author} {\bibinfo {author} {\bibfnamefont {K.}~\bibnamefont
  {Weise}}\ and\ \bibinfo {author} {\bibfnamefont {W.}~\bibnamefont {Wogner}},\
  }\bibfield  {title} {\bibinfo {title} {Comparison of two measurement results
  using the {B}ayesian theory of measurement uncertainty},\ }\href
  {https://doi.org/https://doi.org/10.1088/0957-0233/5/8/001} {\bibfield
  {journal} {\bibinfo  {journal} {Meas. Sci. Technol.}\ }\textbf {\bibinfo
  {volume} {5}},\ \bibinfo {pages} {879} (\bibinfo {year} {1995})}\BibitemShut
  {NoStop}%
\bibitem [{\citenamefont {Zilges}\ \emph {et~al.}(2022)\citenamefont {Zilges},
  \citenamefont {Balabanski}, \citenamefont {Isaak},\ and\ \citenamefont
  {Pietralla}}]{zil22}%
  \BibitemOpen
  \bibfield  {author} {\bibinfo {author} {\bibfnamefont {A.}~\bibnamefont
  {Zilges}}, \bibinfo {author} {\bibfnamefont {D.}~\bibnamefont {Balabanski}},
  \bibinfo {author} {\bibfnamefont {J.}~\bibnamefont {Isaak}},\ and\ \bibinfo
  {author} {\bibfnamefont {N.}~\bibnamefont {Pietralla}},\ }\bibfield  {title}
  {\bibinfo {title} {Photonuclear reactions—from basic research to
  applications},\ }\href
  {https://doi.org/https://doi.org/10.1016/j.ppnp.2021.103903} {\bibfield
  {journal} {\bibinfo  {journal} {Prog. Part. Nucl. Phys.}\ }\textbf {\bibinfo
  {volume} {122}},\ \bibinfo {pages} {103903} (\bibinfo {year}
  {2022})}\BibitemShut {NoStop}%
\bibitem [{\citenamefont {Amano}\ \emph {et~al.}(2009)\citenamefont {Amano},
  \citenamefont {Horikawa}, \citenamefont {Ishihara}, \citenamefont {Miyamoto},
  \citenamefont {Hayakawa}, \citenamefont {Shizuma},\ and\ \citenamefont
  {Mochizuki}}]{ama09}%
  \BibitemOpen
  \bibfield  {author} {\bibinfo {author} {\bibfnamefont {S.}~\bibnamefont
  {Amano}}, \bibinfo {author} {\bibfnamefont {K.}~\bibnamefont {Horikawa}},
  \bibinfo {author} {\bibfnamefont {K.}~\bibnamefont {Ishihara}}, \bibinfo
  {author} {\bibfnamefont {S.}~\bibnamefont {Miyamoto}}, \bibinfo {author}
  {\bibfnamefont {T.}~\bibnamefont {Hayakawa}}, \bibinfo {author}
  {\bibfnamefont {T.}~\bibnamefont {Shizuma}},\ and\ \bibinfo {author}
  {\bibfnamefont {T.}~\bibnamefont {Mochizuki}},\ }\bibfield  {title} {\bibinfo
  {title} {Several-{M}e{V} $\gamma$-ray generation at {N}ew{SUBARU} by laser
  {C}ompton backscattering},\ }\href
  {https://doi.org/https://doi.org/10.1016/j.nima.2009.01.010} {\bibfield
  {journal} {\bibinfo  {journal} {Nucl. Instrum. Methods Phys. Res., Sect. A}\
  }\textbf {\bibinfo {volume} {602}},\ \bibinfo {pages} {337} (\bibinfo {year}
  {2009})}\BibitemShut {NoStop}%
\bibitem [{\citenamefont {Weller}\ \emph {et~al.}(2009)\citenamefont {Weller},
  \citenamefont {Ahmed}, \citenamefont {Gao}, \citenamefont {Tornow},
  \citenamefont {Wu}, \citenamefont {Gai},\ and\ \citenamefont
  {Miskimen}}]{wel09}%
  \BibitemOpen
  \bibfield  {author} {\bibinfo {author} {\bibfnamefont {H.~R.}\ \bibnamefont
  {Weller}}, \bibinfo {author} {\bibfnamefont {M.~W.}\ \bibnamefont {Ahmed}},
  \bibinfo {author} {\bibfnamefont {H.}~\bibnamefont {Gao}}, \bibinfo {author}
  {\bibfnamefont {W.}~\bibnamefont {Tornow}}, \bibinfo {author} {\bibfnamefont
  {Y.~K.}\ \bibnamefont {Wu}}, \bibinfo {author} {\bibfnamefont
  {M.}~\bibnamefont {Gai}},\ and\ \bibinfo {author} {\bibfnamefont
  {R.}~\bibnamefont {Miskimen}},\ }\bibfield  {title} {\bibinfo {title}
  {Research opportunities at the upgraded {HI}$\gamma${S} facility},\ }\href
  {https://doi.org/https://doi.org/10.1016/j.ppnp.2008.07.001} {\bibfield
  {journal} {\bibinfo  {journal} {Prog. Part. Nucl. Phys.}\ }\textbf {\bibinfo
  {volume} {62}},\ \bibinfo {pages} {257} (\bibinfo {year} {2009})}\BibitemShut
  {NoStop}%
\bibitem [{\citenamefont {Fayache}\ \emph {et~al.}(1997)\citenamefont
  {Fayache}, \citenamefont {{von Neumann-Cosel}}, \citenamefont {Richter},
  \citenamefont {Sharon},\ and\ \citenamefont {Zamick}}]{fay97}%
  \BibitemOpen
  \bibfield  {author} {\bibinfo {author} {\bibfnamefont {M.}~\bibnamefont
  {Fayache}}, \bibinfo {author} {\bibfnamefont {P.}~\bibnamefont {{von
  Neumann-Cosel}}}, \bibinfo {author} {\bibfnamefont {A.}~\bibnamefont
  {Richter}}, \bibinfo {author} {\bibfnamefont {Y.}~\bibnamefont {Sharon}},\
  and\ \bibinfo {author} {\bibfnamefont {L.}~\bibnamefont {Zamick}},\
  }\bibfield  {title} {\bibinfo {title} {Effects of the spin-orbit and tensor
  interactions on ${M}1$ excitations in light nuclei},\ }\href
  {https://doi.org/https://doi.org/10.1016/S0375-9474(97)00403-X} {\bibfield
  {journal} {\bibinfo  {journal} {Nucl. Phys. A}\ }\textbf {\bibinfo {volume}
  {627}},\ \bibinfo {pages} {14} (\bibinfo {year} {1997})}\BibitemShut
  {NoStop}%
\bibitem [{\citenamefont {Honma}\ \emph {et~al.}(2004)\citenamefont {Honma},
  \citenamefont {Otsuka}, \citenamefont {Brown},\ and\ \citenamefont
  {Mizusaki}}]{hon04}%
  \BibitemOpen
  \bibfield  {author} {\bibinfo {author} {\bibfnamefont {M.}~\bibnamefont
  {Honma}}, \bibinfo {author} {\bibfnamefont {T.}~\bibnamefont {Otsuka}},
  \bibinfo {author} {\bibfnamefont {B.~A.}\ \bibnamefont {Brown}},\ and\
  \bibinfo {author} {\bibfnamefont {T.}~\bibnamefont {Mizusaki}},\ }\bibfield
  {title} {\bibinfo {title} {New effective interaction for $pf$-shell nuclei
  and its implications for the stability of the ${N}={Z}=28$ closed core},\
  }\href {https://doi.org/10.1103/PhysRevC.69.034335} {\bibfield  {journal}
  {\bibinfo  {journal} {Phys. Rev. C}\ }\textbf {\bibinfo {volume} {69}},\
  \bibinfo {pages} {034335} (\bibinfo {year} {2004})}\BibitemShut {NoStop}%
\bibitem [{\citenamefont {Honma}\ \emph {et~al.}(2005)\citenamefont {Honma},
  \citenamefont {Otsuka}, \citenamefont {Brown},\ and\ \citenamefont
  {Mizusaki}}]{hon05}%
  \BibitemOpen
  \bibfield  {author} {\bibinfo {author} {\bibfnamefont {M.}~\bibnamefont
  {Honma}}, \bibinfo {author} {\bibfnamefont {T.}~\bibnamefont {Otsuka}},
  \bibinfo {author} {\bibfnamefont {B.~A.}\ \bibnamefont {Brown}},\ and\
  \bibinfo {author} {\bibfnamefont {T.}~\bibnamefont {Mizusaki}},\ }\bibfield
  {title} {\bibinfo {title} {Shell-model description of neutron-rich $pf$-shell
  nuclei with a new effective interaction {GXPF}1},\ }\href
  {https://doi.org/10.1140/epjad/i2005-06-032-2} {\bibfield  {journal}
  {\bibinfo  {journal} {Eur. Phys. J. A}\ }\textbf {\bibinfo {volume} {25}},\
  \bibinfo {pages} {499} (\bibinfo {year} {2005})}\BibitemShut {NoStop}%
\bibitem [{\citenamefont {Poves}\ \emph {et~al.}(2001)\citenamefont {Poves},
  \citenamefont {S\'anchez-Solano}, \citenamefont {Caurier},\ and\
  \citenamefont {Nowacki}}]{pov01}%
  \BibitemOpen
  \bibfield  {author} {\bibinfo {author} {\bibfnamefont {A.}~\bibnamefont
  {Poves}}, \bibinfo {author} {\bibfnamefont {J.}~\bibnamefont
  {S\'anchez-Solano}}, \bibinfo {author} {\bibfnamefont {E.}~\bibnamefont
  {Caurier}},\ and\ \bibinfo {author} {\bibfnamefont {F.}~\bibnamefont
  {Nowacki}},\ }\bibfield  {title} {\bibinfo {title} {Shell model study of the
  isobaric chains ${A}=50$, ${A}=51$ and ${A}=52$},\ }\href
  {https://doi.org/https://doi.org/10.1016/S0375-9474(01)00967-8} {\bibfield
  {journal} {\bibinfo  {journal} {Nucl. Phys. A}\ }\textbf {\bibinfo {volume}
  {694}},\ \bibinfo {pages} {157} (\bibinfo {year} {2001})}\BibitemShut
  {NoStop}%
\bibitem [{\citenamefont {Caurier}\ \emph {et~al.}(2005)\citenamefont
  {Caurier}, \citenamefont {Martínez-Pinedo}, \citenamefont {Nowacki},
  \citenamefont {Poves},\ and\ \citenamefont {Zuker}}]{caurier_shell_2005}%
  \BibitemOpen
  \bibfield  {author} {\bibinfo {author} {\bibfnamefont {E.}~\bibnamefont
  {Caurier}}, \bibinfo {author} {\bibfnamefont {G.}~\bibnamefont
  {Martínez-Pinedo}}, \bibinfo {author} {\bibfnamefont {F.}~\bibnamefont
  {Nowacki}}, \bibinfo {author} {\bibfnamefont {A.}~\bibnamefont {Poves}},\
  and\ \bibinfo {author} {\bibfnamefont {A.~P.}\ \bibnamefont {Zuker}},\
  }\bibfield  {title} {\bibinfo {title} {The shell model as a unified view of
  nuclear structure},\ }\href {https://doi.org/10.1103/RevModPhys.77.427}
  {\bibfield  {journal} {\bibinfo  {journal} {Rev. Mod. Phys.}\ }\textbf
  {\bibinfo {volume} {77}},\ \bibinfo {pages} {427} (\bibinfo {year}
  {2005})}\BibitemShut {NoStop}%
\end{thebibliography}%

\end{document}